\documentclass[12pt,manuscript]{aastex}
\usepackage[graphicx]{}
\usepackage{multirow}

\newcommand{\vy}[2]{#1_{\scriptscriptstyle #2}}

\def\gtorder{\mathrel{\raise.3ex\hbox{$>$}\mkern-14mu
             \lower0.6ex\hbox{$\sim$}}}
\def\ltorder{\mathrel{\raise.3ex\hbox{$<$}\mkern-14mu
             \lower0.6ex\hbox{$\sim$}}}
\def\proptwid{\mathrel{\raise.3ex\hbox{$\propto$}\mkern-14mu
             \lower0.6ex\hbox{$\sim$}}}
\def\0946{PG~0946+301}

\def\arcsec{\ifmmode '' \else $''$\fi}

\def\arcsecpoint{\ifmmode ''\!. \else $''\!.$\fi}

\def\kms{\ifmmode {\rm km\ s}^{-1} \else km s$^{-1}$\fi}
\def\Msun{\ifmmode {\rm M}_{\odot} \else M$_{\odot}$\fi}
\def\Lsun{\ifmmode {\rm L}_{\odot} \else L$_{\odot}$\fi}
\def\Zsun{\ifmmode {\rm Z}_{\odot} \else Z$_{\odot}$\fi}

\def\ergscm2{ergs\,s$^{-1}$\,cm$^{-2}$}
\def\icm3{{\rm cm}^{-3}}
\def\icm2{{\rm cm}^{-2}}
\def\qo{\ifmmode q_{\rm o} \else $q_{\rm o}$\fi}
\def\Ho{\ifmmode H_{\rm o} \else $H_{\rm o}$\fi}
\def\ho{\ifmmode h_{\rm o} \else $h_{\rm o}$\fi}
\def\ltsim{\raisebox{-.5ex}{$\;\stackrel{<}{\sim}\;$}}
\def\gtsim{\raisebox{-.5ex}{$\;\stackrel{>}{\sim}\;$}}
\def\vFWHM{\ifmmode v_{\mbox{\tiny FWHM}} \else
            $v_{\mbox{\tiny FWHM}}$\fi}
\def\CCF{\ifmmode F_{\it CCF} \else $F_{\it CCF}$\fi}
\def\ACF{\ifmmode F_{\it ACF} \else $F_{\it ACF}$\fi}
\def\Halpha{\ifmmode {\rm H}\alpha \else H$\alpha$\fi}
\def\Hbeta{\ifmmode {\rm H}\beta \else H$\beta$\fi}
\def\Hgamma{\ifmmode {\rm H}\gamma \else H$\gamma$\fi}
\def\Hdelta{\ifmmode {\rm H}\delta \else H$\delta$\fi}
\def\Lya{\ifmmode {\rm Ly}\alpha \else Ly$\alpha$\fi}
\def\Lyb{\ifmmode {\rm Ly}\beta \else Ly$\beta$\fi}
\def\Lyg{\ifmmode {\rm Ly}\beta \else Ly$\gamma$\fi}
\def\hi{H\,{\sc i}}

\def\cii{C\,{\sc ii}}
\def\ciii{\ifmmode {\rm C}\,{\sc iii} \else C\,{\sc iii}\fi}
\def\civ{\ifmmode {\rm C}\,{\sc iv} \else C\,{\sc iv}\fi}
\def\ni{N\,{\sc i}}

\def\nv{N\,{\sc v}}
\def\oi{O\,{\sc i}}

\def\o5007{[O\,{\sc iii}]\,$\lambda5007$}
\def\oiv{O\,{\sc iv}}

\def\ovi{O\,{\sc vi}}

\def\siiv{Si\,{\sc iv}}
\def\siIV{Si\,{\sc iv}}
\def\siIII{Si\,{\sc iii}}
\def\siII{Si\,{\sc ii}}

\def\sii{S\,{\sc ii}}

\def\siv{S\,{\sc iv}}

\def\feii{Fe\,{\sc ii}}

\def\o{\o}
\begin{document}

\title{A 10 kpc Scale Seyfert Galaxy Outflow:\\
 HST/COS Observations of IRAS F22456-5125}


\author{
 Benoit C.J. Borguet\altaffilmark{1},
 Doug Edmonds\altaffilmark{1},
 Nahum Arav\altaffilmark{1}, 
 Jay Dunn\altaffilmark{2},
 Gerard A. Kriss\altaffilmark{3,4}     
}

\altaffiltext{1}{Department of Physics, Virginia Tech, Blacksburg, Va 24061; email: benbo@vt.edu}
\altaffiltext{2}{Augusta Perimeter College, Atlanta, GA}
\altaffiltext{3}{Space Telescope Science Institute, 3700 San Martin Drive, Baltimore, MD 21218, USA}
\altaffiltext{4}{Center for Astrophysical Sciences, Department of Physics and Astronomy, Johns Hopkins University, Baltimore, MD 21218}

\begin{abstract}

We present analysis of the UV-spectrum of the low-$z$ AGN
IRAS-F22456-5125 obtained with the Cosmic Origins Spectrograph on
board the Hubble Space Telescope. The spectrum reveals six main
kinematic components, spanning a range of velocities of up to 800 km
s$^{-1}$, which for the first time are observed in troughs associated
with \cii, \civ, \nv, \siII, \siIII, \siiv\ and \siv. We also obtain data
on the \ovi\ troughs, which we compare to those available from an earlier
FUSE epoch. Column densities measured from these ions allow us to derive a well-constrained
photoionization solution for each outflow component.  Two of these
kinematic components show troughs associated with transitions from
excited states of \siII\ and \cii. The number density inferred from
these troughs, in combination with the deduced ioinization parameter,
allows us to determine the distance to these outflow components from
the central source. We find these components to be at a distance of $\sim$ 10 kpc.
The distances and the number densities derived are consistent with the outflow being part of a galactic
wind.

\end{abstract}

\keywords{galaxies: quasars ---
galaxies: individual (IRAS F22456-5125) ---
line: formation ---
quasars: absorption lines}

\section{INTRODUCTION}

Mass outflows are detected in the UV spectra of more than 50\% of low
redshift active galactic nuclei (AGN) mainly Seyfert galaxies,
e.g.~\citet{Crenshaw99},~\citet{Kriss02},~\citet{Dunn07},~\citet{Ganguly08}.
These outflows are observed as narrow absorption lines (a few hundred km
s$^{-1}$ in width) blueshifted with respect to the AGN systemic
redshift.

In this paper, we determine the ionization equilibrium, distance, mass flow rate, and kinetic luminosity of the UV outflow observed in the luminous Seyfert
1 galaxy IRAS F22456-5125 ($z=0.1016$, \citealt{Dunn10b}). The bolometric luminosity of this object, $L_{bol} = 10^{45.6}$ ergs s$^{-1}$ (see Section~\ref{analiab}), places it at the
Seyfert/quasar border defined to be $10^{12} L_{\sun}$, where $L_{\sun}$ is the luminosity of the sun \citep{Soifer87}.
Several absorption systems are resolved in the UV spectrum in five main kinematic components ranging in velocities from 
$-$20 km s$^{-1}$ to $-$820 km s$^{-1}$. A detailed analysis of the physical properties of the UV absorber determined from Far
Ultraviolet Spectroscopic Explorer (FUSE) archival spectra has been published by~\citet{Dunn10b}. These authors
report a lower limit on the distance $R$ of the absorbing material from the central source of 20 kpc using photoionization timescale arguments.

In June 2010 we observed IRAS F22456-5125 with the Cosmic Origins Spectrograph (COS) on
board the Hubble Space Telescope (HST) as part of our program aiming at determining the cosmological impact of AGN outflows
(PI: Arav, PID: 11686). The high signal to noise spectrum obtained reveals the presence of absorption troughs associated
with high ionization species (\civ, \nv, \ovi, \siiv\ and \siv) as well as lower ones (\siII, \siIII, \cii) thus increasing the number of constraints
on the photoionization analysis of the absorber compared to~\citet{Dunn10b}. We also identify absorption troughs corresponding to excited states
of \siII\ and \cii\ associated with two kinematic components of the UV outflow. The population of the excited state 
relative to the resonance counterpart provides an indirect measurement of the number density of the gas producing the lines~\citep{AGN^3}.
These number densities allow us to determine reliable distances to these two components and hence derive their mass flow rates and kinetic luminosities.

The plan of the paper is as follows: in \S~\ref{dataredu} we present the COS observations of IRAS F22456-5125 as well as the reduction of the data
and identification of the spectral features within the COS range. In \S~\ref{coldensisec} we detail the computation
of the column densities associated with every species. We present the photoionization analysis of the outflow
components in \S~\ref{analiab} and report the derived distance, mass flow rate, and kinetic luminosity in \S~\ref{distancesect}. We conclude
the paper by a discussion of our results in \S~\ref{discussion}. This paper is the second of a series and the reader will be
referred to~\citet[hereafter Paper I]{Edmonds11} for further details on the techniques used throughout the paper.

\section{HST/COS observations and data reduction}
\label{dataredu}

We observed IRAS F22456-5125 using the COS instrument~\citep{Osterman10} on board the HST in June 2010
using both medium resolution ($\Delta \lambda / \lambda \sim 18,000$) Far Ultraviolet
gratings G130M and G160M. Sub-exposures of the target were obtained for each grating
through the Primary Science Aperture (PSA) using different central wavelength settings in
order to minimize the impact of the instrumental features as well as to fill the gap
between detector segments providing a continuous coverage
over the spectral range between roughly 1135-1795 \AA. We obtained a total integration
time of 15,056 s and 11,889 s for the G130M and G160M gratings, respectively.

\begin{figure}
  \includegraphics[angle=90,width=1.0\textwidth]{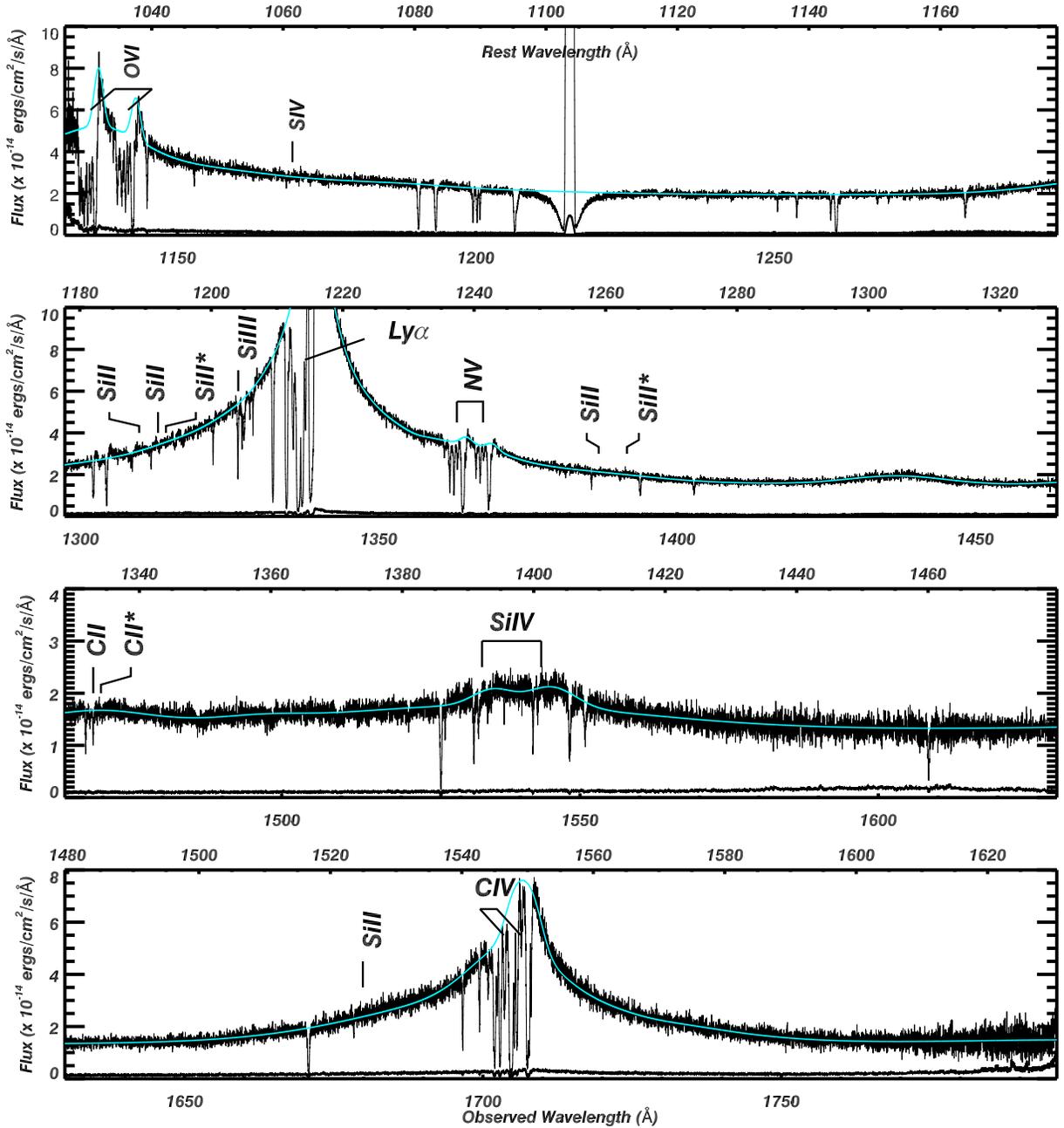}\\
 \caption{The full FUV spectrum of IRAS F22456-5125
          obtained by COS. The major absorption troughs related
          to the intrinsic absorber are labeled. A full identification of all
          the absorption features is presented in the online version. The green line
          represent our fit to the non absorbed emission model (see Section~\ref{emimo}).}
 \label{fullspec} 
\end{figure}

The datasets processed through the standard CALCOS\footnote{Details on CALCOS can
be found in the COS Data Handbook.} pipeline were retrieved from the
MAST archive. They were then flat-fielded and combined together using the COADD\textunderscore X1D\footnote{The
routine can be found at http://casa.colorado.edu/$\sim$danforth/science/cos/costools.html} IDL
pipeline developed by the COS GTO team (see~\citealt{Danforth10} for details).
The reduced spectrum with its original $\sim 2~ \mathrm{km}~ \mathrm{s}^{-1}$ oversampling has an overall signal
to noise ratio $\geq 15$ per pixel in most of the continuum region. Typical errors in the wavelength calibration are
less than $15 ~ \mathrm{km}~ \mathrm{s}^{-1}$. In Figure~\ref{fullspec}, we show the majority
of the spectrum on which we identified major intrinsic absorption features associated with the outflow.
The COS FUV spectrum of IRAS F22456-5125 is presented in greater detail along with the
identification of most absorption features (interstellar, intergalactic, and intrinsic lines)
in the on-line version of Figure~\ref{fullspec}.

\subsection{Identification of spectral features}
\label{identi}

Using archived FUSE spectra~\citet{Dunn07,Dunn10b} reported the first detection of five distinct
kinematic components with centroid velocities $v_1 = -800~ \mathrm{km}~ \mathrm{s}^{-1}$,
$v_2 = -610~ \mathrm{km}~ \mathrm{s}^{-1}$,  $v_3 = -440~ \mathrm{km}~ \mathrm{s}^{-1}$,
$v_4 = -320~ \mathrm{km}~ \mathrm{s}^{-1}$, $v_5 = -130~ \mathrm{km}~ \mathrm{s}^{-1}$, and
FWHM $\in [50, 200]~ \mathrm{km}~ \mathrm{s}^{-1}$ associated with an intrinsic UV outflow
in IRAS F22456-5125. These components, spanning a total velocity range of $800~ \mathrm{km}~ \mathrm{s}^{-1}$,
were detected in \ovi, \ciii\ and in several lines of the Lyman series (Ly$\beta$ to Ly$\eta$).
Using the kinematic pattern reported by~\citet{Dunn10b} as a template we identify absorption
features in our COS spectrum related to both low ionization (\cii, \siII, \siIII)
and high ionization species (\siIV, \siv, \civ, \nv, \ovi) as well as in the Ly$\alpha$ transition.
Absorption troughs from the metastable level \cii* $\lambda 1335.704$ are detected
in components $2$ and $3$, and troughs from metastable \siII* $\lambda 1264.738$ and $\lambda 1194.500$
are detected in component $2$.

While the absorption troughs associated with the higher ionization lines generally exhibit broader
profiles, we observe a 1:1 kinematic correspondence between the core of these components and the narrower features
associated with the lower ionization species of the outflow. Given the significantly broad range of velocities 
covered by the components and their net kinematic separations, such a match is not likely to occur by chance. This argues in
favor of a scenario where the troughs of the different ionic species detected in a given kinematic component are
generated in the same region. This observation is strengthened by the fact that most of the troughs have a line
profile similar to that of the non-blended \nv\ $\lambda 1238.820$ line when properly scaled.

The high S/N of our COS observations (S/N $\gtrsim 40$ per resolution element on most of the spectral coverage) reveals the presence of kinematic substructures in several
components of the outflow compared to the lower S/N FUSE observations (S/N $\sim 7$,~\citealt{Dunn10b}).
Nevertheless given the self-blending of these features in the strongest lines
(e.g. \ovi) and the absence of apparent change between the FUSE and COS observations, we will use
the labeling of components as defined in~\citet{Dunn10b}. We will however
separate their trough 5 into low and high velocity components given the apparent difference
in ionization suggested by the presence of a stronger \siIII\ in subcomponent $5A$ ($v_{5A} = -40~ \mathrm{km}~ \mathrm{s}^{-1}$)
than in $5B$ ($v_{5B} = -130~ \mathrm{km}~ \mathrm{s}^{-1}$) relative to the higher ionization lines (\civ, \nv, \ovi, see Figure~\ref{alllineprof}).
Most of our analysis in this paper concentrates on components 2 and 3 of the outflow, for which
absorption features associated with an excited state have been detected.

\subsection{Deconvolution of the COS spectrum}
\label{decosec}

Detailed analysis of the on-orbit COS Line Spread Function (LSF) revealed the presence of
broadened wings that scatter a significant part of the continuum flux inside the
absorption troughs (see~\citealt{Kriss11} for details). This continuum leaking is particularly
strong for narrow absorption troughs (FWHM $\sim 50~ \mathrm{km}~ \mathrm{s}^{-1}$) in which this effect
may significantly affect the estimation of the true column density by artificially increasing the residual
intensity observed inside the troughs.

\begin{figure}
  \includegraphics[angle=90,width=1.0\textwidth]{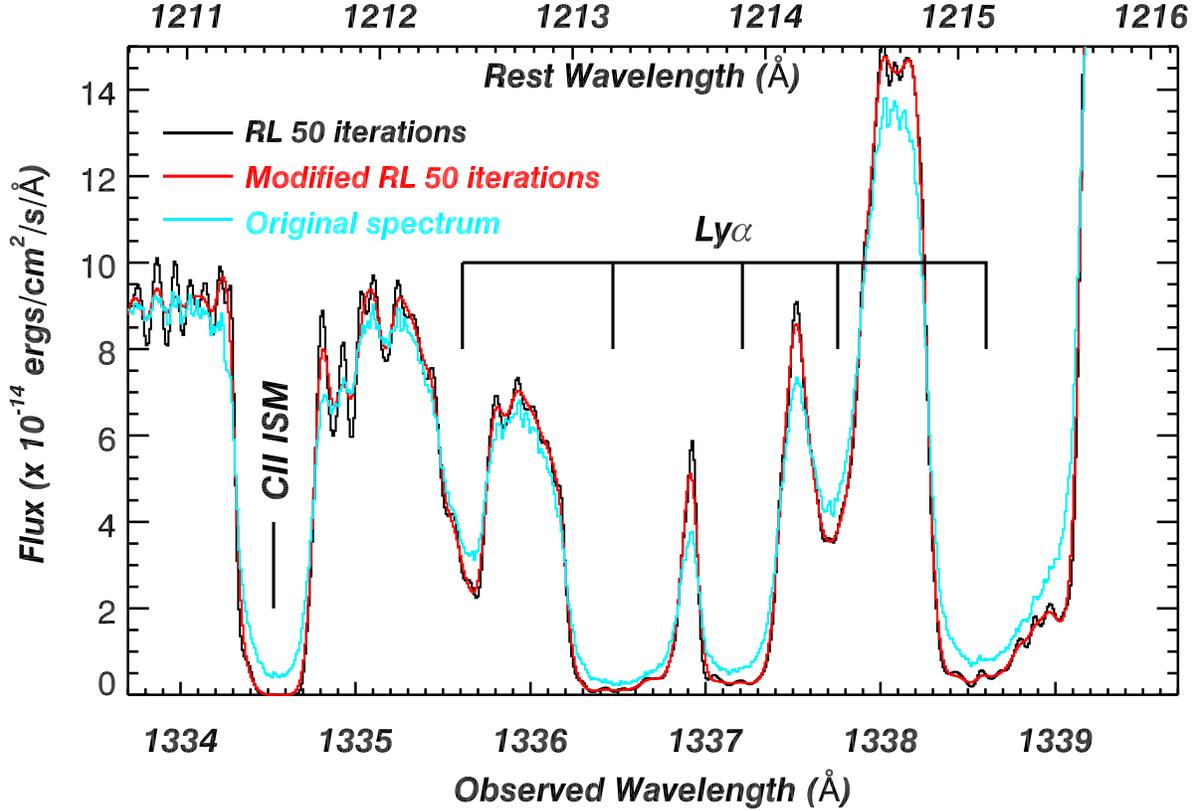}\\
 \caption{Illustration of the necessity of using a deconvolution algorithm when dealing with COS
data (see text for details). Troughs associated with the intrinsic absorber significantly deepen
while the saturated interstellar \cii\ line exhibits
the expected squared black bottom profile. The main difference between the deconvolved spectrum
using the RL method and the modified RL algorithm respecting the sampling theorem is the
significant reduction of oscillations due to the total deconvolution process performed in RL,
even more so when considering a high number of iterations.}
 \label{deconv} 
\end{figure}

Given the overall good signal to noise ratio of our data, we can correct the effect
of the poor LSF by deconvolving the COS spectrum. Adopting the procedure described in \citet{Kriss11}, we deconvolve the spectrum obtained for each grating
in 50 \AA~ intervals using the wavelength dependent LSFs 
and an IDL implementation of the stsdas Richardson-Lucy ``lucy'' algorithm (G. Schneider \& B.
Stobie, private communication, 2011). The main effect of the deconvolution is illustrated in Figure~\ref{deconv} in which we
clearly see that the deconvolved spectrum shows significantly deepened intrinsic Ly$\alpha$ absorption troughs
as well as produces a square, black bottom for the saturated interstellar line \cii\ $\lambda 1334.532$.

However, the main drawback of the deconvolution algorithms commonly used, such as the
Richardson-Lucy (RL) algorithm, is a significant increase of the noise in the deconvolved
spectrum due to the fact that these techniques try to perform a total deconvolution of the data,
i.e. in which the LSF of the deconvolved spectrum is a Dirac delta function, violating the Shannon sampling
theorem (see~\citealt{Magain98} for a thorough discussion of these issues). In order to decrease these
effects, we modified the RL algorithm by forcing the deconvolved
spectrum to have a LSF satisfying the sampling theorem. We choose the deconvolved LSF to be a
Gaussian with a 2 pixel FWHM ($\sim 5~ \mathrm{km}~ \mathrm{s}^{-1}$ given the COS detector sampling). This operation prevents
the appearance of strong unwanted oscillations since we force the maximum resolution that can be achieved
in the deconvolved data to agree with the sampling theorem. The deconvolved spectrum produced by
this modified Richardson Lucy algorithm is similar to the one obtained by the traditional RL algorithm (see Figure~\ref{deconv}), the main
difference being the significant decrease of the high frequencies and high amplitude features
artificially generated by RL deconvolution with a high number of iterations.
In our analysis, we will derive the column density for each ionic species using the spectrum deconvolved
with the modified RL algorithm, allowing us to derive more accurate
column densities associated with the narrow absorption components observed in IRAS F22456-5125.

\subsection{Unabsorbed emission model}
\label{emimo}

The unabsorbed emission model $F_0(\lambda)$ of IRAS F22456-5125 is constructed in a similar manner to the one described in detail
for IRAS-F04250-5718 in Paper I, in which we consider three main sources of emission : a continuum,
a broad emission line (BEL) component and a narrow emission line (NEL) component. Adopting a single power law
$F(\lambda) = F_{1150}(\lambda/1150)^{\alpha}$ to describe the deredenned (E(B-V)=0.01035, \citealt{Schlegel98})
continuum emission, we obtain a reduced $\chi^{2}_{red} = 1.413$ over emission/absorption line free regions of the rest wavelength spectrum ([1115, 1130] \AA, [1340,1360] \AA, and [1455,1475] \AA)
with $\alpha=-1.473 \pm 0.068$ and $F_{1150}=2.130~ 10^{-14} \pm 0.0033~ 10^{-14}$ erg cm$^{-2}$\AA$^{-1}$s$^{-1}$.

Prominent BEL features observed in the spectrum (Ly$\alpha$, \civ, \ovi)
are fit using two broad gaussians of FWHM $\sim 9000$ and $2000 ~ \mathrm{km}~ \mathrm{s}^{-1}$. The NEL 
component of each line of a doublet is fit by a single narrower gaussian (FWHM $\sim 600 ~ \mathrm{km}~ \mathrm{s}^{-1}$)
centered around the rest wavelength of each line, with the separation of the two
gaussians fixed to the velocity difference between the doublet lines. The NEL of the strong Ly$\alpha$
line is best fit by two gaussians of FWHM $\sim 1200$ and $400 ~ \mathrm{km}~ \mathrm{s}^{-1}$.
The remaining weaker emission features in the spectrum (\siiv+\oiv, \cii, \nv, \oi\ etc.) are
modeled by a smooth cubic spline fit. A normalized spectrum is then obtained by dividing the data with the emission model.
We present our best fit to the unabsorbed spectrum of IRAS F22456-5125 in Figure~\ref{fullspec}

\section{Column Density Determination}
\label{coldensisec}

\subsection{Methodology}

The column density associated with a given ionic species detected in the outflow is
determined by modeling the residual intensity in the normalized data of the absorption
troughs. Assuming a single homogeneous emission source $F_0(v)$ and a one-dimensional spatial distribution of
optical depth across the emission source $\tau_i(x,v)$, we can express the intensity $F_i(v)$
observed for a line $i$ as~\citep{Arav05}:
\begin{equation}
  F_i(v) = F_0(v) \int_{0}^{1} e^{-\tau_i(x,v)} dx 
\end{equation}
where $v$ is the radial velocity of the outflow, and the spatial extension of the emission source is
normalized to 1. Once the optical depth solution $\tau_i(x,v)$ is derived at a given radial velocity, we
link the observed residual intensity $I_i(v) = F_i(v)/ F_0(v)$ to the ionic column density using the relation :
\begin{equation}
 N_{ion}(v) =  \frac{3.8 \times 10^{14}}{f_i \lambda_i} < \tau_i(v) > ~ (\mathrm{cm}^{-2} \mathrm{km}^{-1}~\mathrm{s})
\end{equation}
where $f_i$, $\lambda_i$, and $< \tau_i(v) >$ are the oscillator strength, the
rest wavelength, and the average optical depth across the emission source of line $i$ (see Paper I), respectively.

We consider here the three absorber models (i.e. optical depth distributions) discussed in Paper I;
the Apparent Optical Depth (AOD), Partial Covering (PC), and Power Law (PL) models.
We use these three models in order to account for possible inhomogeneities in the absorber (see Sect.~\ref{discussion}),
which cause the apparent strength ratio $R_a = \tau_{i} /\tau_{j}$ of two
lines $i,~j$ from a given ion to deviate from the expected laboratory ratio $R_l = \lambda_i f_i / \lambda_j f_j$
(e.g.~\citealt{Wampler95},~\citealt{Hamann97},~\citealt{Arav99}). Wherever possible we derive these three optical
depth solutions for ions with multiple transitions. However, as mentioned in Paper I, we consider the results obtained with the PL model performed on doublets with caution
given its increased sensitivity to the S/N, which can lead to severe overestimation
of the underlying column density ~\citep{Arav05}. For singlet lines we will generally only derive
a lower limit on the column density using the AOD method. This lower limit
will be considered a measurement in cases where the singlet line is associated
with a kinematic component for which other multiplets do not show signs of saturation.
In the following subsections, we use the term (non-black) saturation to qualify 
absorption troughs of doublets in which $R_a = \tau_{i} /\tau_{j} < 0.75 ~ R_l $,
where $\tau_i$  and $\tau_j$ are the apparent optical depth of the strongest and
the weakest component of the doublet, respectively.

\subsection{Column density measurements}

\begin{figure}
  \includegraphics[angle=90,width=1.0\textwidth]{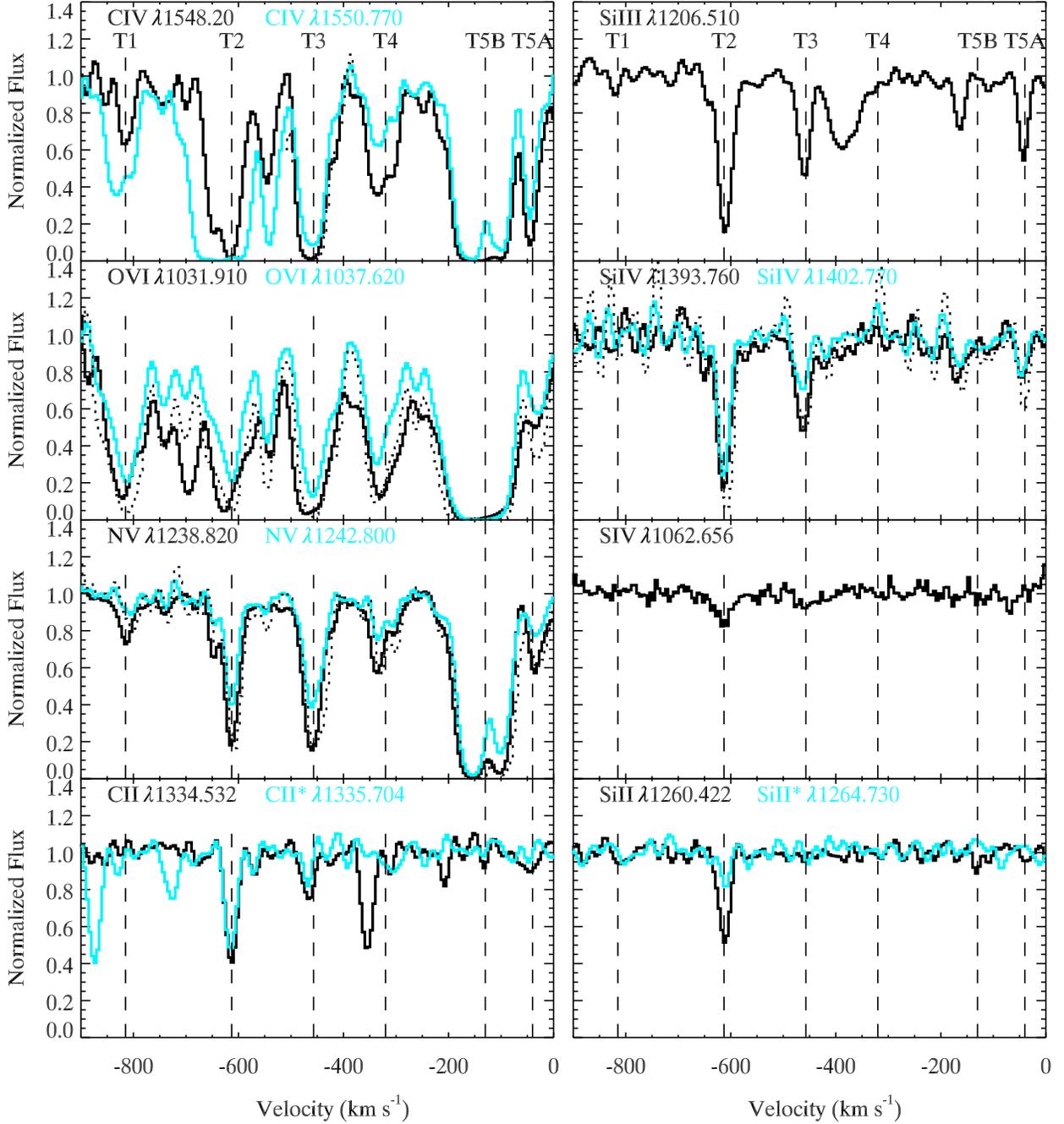}\\
 \caption{Normalized absorption line profile of the metal lines associated with the
outflow in IRAS F22456-5125. The line profiles have been deconvolved using the modified
RL algorithm described in Section~\ref{decosec} and rebinned to a common $\sim 5~ \mathrm{km}~ \mathrm{s}^{-1}$
dispersion velocity scale. For doublets, we overplot the expected residual intensity in the strongest
component based on the residual flux observed in the weakest component assuming an AOD absorber model.
For \civ\ we only plot that quantity in regions free of self blending (mainly $T3$, see text).}
 \label{alllineprof} 
\end{figure}

Computed ionic column densities are determined using the deconvolved line profiles
presented in Figure~\ref{alllineprof} and the ionic transition properties reported in
Table~\ref{line_data}. The computed column densities are reported in Table~\ref{coldensi}
for the three absorber models when possible. The adopted values shown in the last column
of Table~\ref{coldensi} are the ones used in the photoionization
analysis. When available, we choose to use the value reported in the PC column as the measurement and use the
PL measurement and error as the upper error in order to account for the possible inhomogeneities in the absorbing material distribution. If only the AOD
determination is available we will consider the reported value minus the error as a lower limit unless we have evidence suggesting
a high covering.

 \begin{deluxetable}{lrlll}
\tabletypesize{\footnotesize}
\tablecaption{{\sc Atomic Data for the Observed Transitions}}
\tablewidth{0pt}
\tablehead{
\colhead{Ion}
&\colhead{$E_{\mathrm{low}}^{\mathrm{a}}$}
&\colhead{$\lambda_i^{\mathrm{b}}$}
&\colhead{$g_{\mathrm{low}}^{\mathrm{c}}$}
&\colhead{$f_i^{\mathrm{d}}$}\\
\colhead{}
&\colhead{cm$^{-1}$}
&\colhead{\AA}
&\colhead{}
&\colhead{}

}

\startdata

\hi         &  0.00   &   1215.670  &     2    &  0.4164  \\

\cii        &  0.00   &   1334.532  &     2    &  0.1290  \\
\cii*       &  63.42  &   1335.704$^{\mathrm{e}}$  &     4    &  0.1277  \\

\civ        &  0.00   &   1548.202  &     2    &  0.1900   \\
\civ        &  0.00   &   1550.774  &     2    &  0.0952   \\

\nv         &  0.00   &   1238.821  &     2    &  0.1560   \\
\nv         &  0.00   &   1242.804  &     2    &  0.0780   \\

\ovi        &  0.00   &   1031.912  &     2    &  0.1330   \\
\ovi        &  0.00   &   1037.613  &     2    &  0.0660   \\

\siII       &  0.00   &   1190.416 &      2    &  0.2770   \\
\siII       &  0.00   &   1193.280 &      2    &  0.5750   \\
\siII*      &  287.24 &   1194.500 &      4    &  0.7370   \\
\siII       &  0.00   &   1260.422 &      2    &  1.2200   \\
\siII*      &  287.24 &   1264.730 &      4    &  1.0900   \\
\siII       &  0.00   &   1304.370 &      2    &  0.0928   \\
\siII       &  0.00   &   1526.720 &      2    &  0.1330   \\

\siIII      &  0.00   &   1206.500 &      1    &  1.6700   \\

\siIV       &  0.00   &   1393.760 &      2    &  0.5130   \\
\siIV       &  0.00   &   1402.770 &      2    &  0.2550   \\

\siv        &  0.00   &   1062.656 &      2    &  0.0500  \\

\enddata
\label{line_data}
a - Lower level energy.
b - Wavelength of the transition.
c - Statistical weight.
d - Oscillator strength. We use the oscillator strengths from the National Institute of Standards and Technology
(NIST) database, except for \siv\ for which we use the value reported in~\citet{Hibbert02}.
e - Blend of two transitions, we report the sum of the oscillator strength and the weighted average of $\lambda_i$. 
\end{deluxetable}

 \begin{deluxetable}{crlrrrr}

\tabletypesize{\tiny}
\tablecaption{{\sc Computed column densities}}
\tablewidth{0pt}
\tablehead{
\colhead{Trough}
&\colhead{$v_i$}
&\colhead{Ion}
&\colhead{AOD$^{\mathrm{a}}$}
&\colhead{PC$^{\mathrm{a}}$}
&\colhead{PL$^{\mathrm{a}}$}
&\colhead{Adopted$^{\mathrm{f}}$}\\
\colhead{}
&\colhead{km s$^{-1}$}
&\colhead{}
&\colhead{$10^{12} cm^{-2}$}
&\colhead{$10^{12} cm^{-2}$}
&\colhead{$10^{12} cm^{-2}$}
&\colhead{$10^{12} cm^{-2}$}
}

\startdata

T1  &  -800   &   \hi   &  73.9$^{+0.4}_{-0.4}$    &              $< 900^{\mathrm{b}}$ &           &   $\in [$73.5, 900$]$     \\
    &     &   \civ  &  20.9$^{+1.1}_{-1.0}$    &       ...                 &        ...                   &   20.9$^{+1.1}_{-1.0}$      \\
    &     &   \nv   &  13.9$^{+1.6}_{-1.6}$    & 18.1$^{+1.8}_{-1.6}$   &  20.2$^{+1.6}_{-1.2}$    &   18.1$^{+3.7}_{-1.6}$      \\
    &     &  \ovi   &  474$^{+9}_{-9}$         & 745$^{+137}_{-28}$     &    ...                       &   745$^{+137}_{-28}$      \\
    &     &  \siIII & $<$ 0.32$^{+0.04}_{-0.04}$  &         ...                &      ...                  &   $<$ 0.36     \\

T2  &  -610   &   \hi   &  436$^{+56}_{-1}$        & $4400^{+660 ~\mathrm{b}}_{-660}$ &     ...        &   $\in [$9400, 15800$^{\mathrm{e}}]$    \\
    &     &   \cii  &  51.0$^{+2.4}_{-2.2}$    & 59.7$^{+3.3~\mathrm{c}}_{-2.9}$   &      ...          &   $>$ 48.8     \\
    &     &   \cii* &  43.2$^{+2.3}_{-2.1}$    & 49.5$^{+3.2~\mathrm{c}}_{-2.8}$   &      ...          &   $\gtsim$ 41.1      \\
    &     &   \civ  &  251$^{+27}_{-5}$        &       ...                  &             ...              &   $>$ 251    \\
    &     &   \nv   &  109$^{+2.6}_{-2.5}$     & 118$^{+15}_{-2}$       &  142$^{+6.4}_{-3.4}$     &   118$^{+30.4}_{-2}$     \\
    &     &  \ovi   &  604$^{+11}_{-10}$       & 816$^{+127}_{-28}$     & 1199$^{+272}_{-11}$      &   816$^{+655}_{-28}$      \\
    &     & \siII   &  10.5$^{+0.6}_{-0.6}$    & 13.7$^{+1.2}_{-0.9}$   & 33.3$^{+3.5}_{-2.8}$     &   13.7$^{+23.1}_{-0.9}$     \\
    &     & \siII*  &  1.04$^{+0.13}_{-0.12}$  & 1.18$^{+0.18}_{-0.15}$ & 1.59$^{+0.20}_{-0.17}$   &   1.18$^{+0.61}_{-0.15}$     \\
    &     & \siIII &  $>$ 9.24$^{+0.16}_{-0.15}$  &         ...             &       ...                    &   $>$ 9.08     \\
    &     & \siIV  &  36.3$^{+0.2}_{-0.2}$     & 49.4$^{+3.7}_{-2.6}$   &        ...                   &   $>$ 46.8     \\
    &     & \siv   &  54.0$^{+4.5}_{-4.5}$     &                        &        ...                   &   $\gtsim$49.5     \\

T3  &  -440   &   \hi   &  275$^{+2}_{-2}$         & $4230^{+790 ~\mathrm{b}}_{-790}$ &       ...      &   $4230^{+790}_{-790}$   \\
    &     &   \cii  &  19.9$^{+2.4}_{-2.0}$    &          ...               &      ...                     &   $>$ 17.9      \\
    &     &   \cii* &  7.66$^{+1.55}_{-1.27}$  &            ...             &     ...                      &   $>$ 6.39     \\
    &     &   \civ  &  301$^{+5}_{-5}$         & 336$^{+8}_{-6}$        & 432$^{+73}_{-8}$         &   336$^{+169}_{-6}$        \\
    &     &   \nv   &  143$^{+3}_{-3}$         & 167$^{+78}_{-5}$       &         ...                  &   167$^{+78}_{-5}$      \\
    &     &  \ovi   &  552$^{+11}_{-9}$        & 644$^{+17}_{-13}$      &        ...                   &   644$^{+17}_{-13}$      \\
    &     & \siII   &  $<$ 0.66                &           ...              &    ...                       &   $<$ 0.66 \\
    &     & \siIII &  3.88$^{+0.09}_{-0.08}$   &           ...              &    ...                       &   $>$ 3.80      \\
    &     & \siIV  &  13.2$^{+1.1}_{-1.0}$     & 15.9$^{+1.8}_{-1.0}$   & 17.3$^{+1.5}_{-1.0}$     &   15.9$^{+2.9}_{-1.0}$      \\
    &     & \siv   &  $<$ 24.2                 &          ...               &     ...                      &  $<$ 24.2      \\

T4  &  -320   &   \hi   &  87.0$^{+0.4}_{-0.4}$    & $550^{+180 ~\mathrm{b}}_{-180}$ &    ...         &   $550^{+180}_{-180}$          \\
    &     &   \civ  &  62.7$^{+2.0}_{-1.9}$    &           ...              &      ...                     &   62.7$^{+2.0}_{-1.9}$      \\
    &     &   \nv   &  46.0$^{+2.2}_{-2.1}$    & 57.5$^{+4.2}_{-2.8}$   & 67.9$^{+4.8}_{-3.6}$     &   57.5$^{+15.2}_{-2.8}$      \\
    &     &  \ovi   &  335$^{+7}_{-7}$         & 400$^{+10}_{-8}$       &       ...                    &   400$^{+10}_{-8}$       \\

T5B &  -130   &   \hi   &  399$^{+50}_{-1}$        & $6010^{+1200 ~\mathrm{b}}_{-1200}$ &     ...      &   $6010^{+1200}_{-1200}$         \\
    &     &   \civ  &  795$^{+104}_{-11}$      &             ...            &           ...                &   $>$ 784       \\
    &     &   \nv   &  935$^{+80}_{-14}$       & 1035$^{+137}_{-14}$    & 1469$^{+284}_{-7}$       &   1035$^{+718}_{-14}$        \\
    &     &  \ovi   &  $>$ 2608$_{-22}$       &           ...              &         ...                   &   $>$ 2586     \\
    &     & \siIII  &  1.78$^{+0.08}_{-0.08}$  &          ...               &       ...                    &   $>$1.70      \\
    &     & \siIV   &  8.49$^{+1.13}_{-1.01}$  & 13.7$^{+11.7}_{-1.9}$  &       ...                    &   13.7$^{+11.7}_{-1.9}$      \\

T5A &  -40   &   \hi   &  88.2$^{+0.5}_{-0.5}$    &   $6010^{+1200 ~\mathrm{d}}_{-1200}$                     &            ...            &   $\in [$87.7, 7210$]$          \\
    &     &   \civ  &  97.4$^{+2.4}_{-2.3}$    &             ...            &           ...                &   $>$95.1      \\
    &     &   \nv   &  34.7$^{+1.8}_{-1.7}$    & 36.7$^{+1.3}_{-1.3}$   & 40.2$^{+1.1}_{-1.1}$     &   36.7$^{+4.6}_{-1.3}$      \\
    &     &  \ovi   &  $>$ 109$_{-4}$         &            ...             &        ...                    &   $>$ 105      \\
    &     & \siIII  &  2.25$^{+0.06}_{-0.06}$  &           ...              &       ...                    &   $>$2.19       \\
    &     & \siIV   &  5.55$^{+0.72}_{-0.69}$  & 13.7$^{+176.8}_{-2.6}$ &          ...                 &   $>$13.7       \\

\enddata
\label{coldensi}
a) The integrated column densities for the three absorber models. The quoted error arise from photon statistics only and are computed using the technique outlined in~\citet{Gabel05a}.\\
b) Estimates from~\citet{Dunn10b}.
c) Using the covering solution of \siIV (see text).
d) \citet{Dunn10b} do not make the distinction between the two sub-components in trough $T5$, so we report an identical PC value in the shallower trough
$T5A$ to be considered as a conservative upper limit since the bulk of the column density is coming from $T5B$.\\
e) The lower is fixed by the detection of Ly10 associated with that component (see text) and the upper limit is given by the absence of a \hi\ bound free edge.
f) Adopted values for the photoionization study (see text).
\end{deluxetable}

\subsubsection{\rm{\hi}}
\label{hicold}

  The spectral coverage of the COS G130M and G160M gratings only allows us to
cover the Ly$\alpha$ line that shows a deep and smooth profile in which the different
kinematic components blend. The absence of higher-order Lyman series lines restrict us
to put a lower limit on the \hi\ column density by applying the AOD method to the
Ly$\alpha$ profile. 

\begin{figure}
  \includegraphics[angle=90,width=1.0\textwidth]{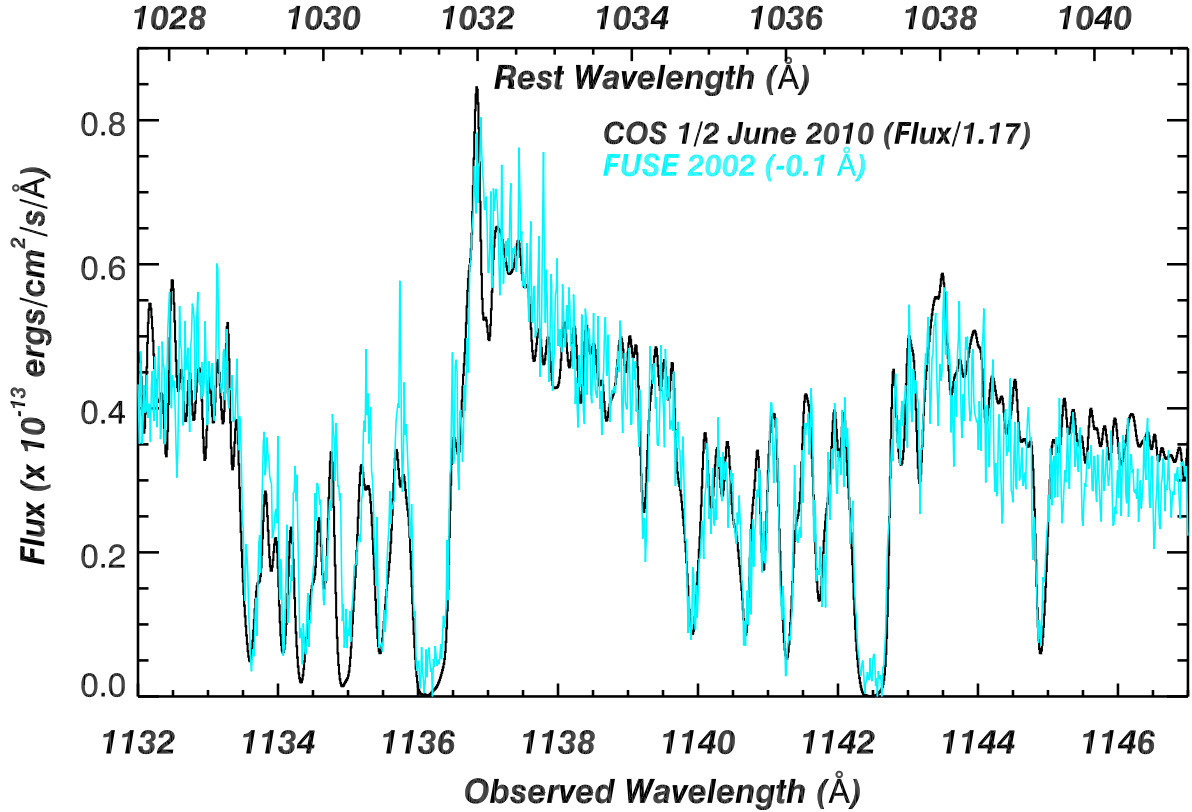}\\
 \caption{Comparison between the higher S/N FUSE 2002 and the deconvolved COS 2010 IRAS F22456-5125 spectra in
 the rest frame \ovi\ region. While the continuum level in 2010 is 17\% higher than in the 2002 FUSE
observation, the absorption trough profile in the red component do not show significant changes given the limited S/N of the FUSE
observation and the larger aperture used by the latter. Changes in the blue line profile seems to be
observed in the blue component of the \ovi\ doublet, however, the line profile is located at their edge of the COS detector 
in a region where the S/N is lower and the wavelength solution is inaccurate.}
 \label{cosfuse_compa} 
\end{figure}

A better constraint on the \hi\ column density is determined by using higher-order Ly-series lines from
earlier FUSE data~\citet{Dunn10b}. In Figure~\ref{cosfuse_compa} we compare
the June 2010 HST/COS and 2004 FUSE spectra of IRAS F22456-5125 in the overlapping region, essentially showing
the rest frame \ovi\ region of the spectrum. While we observe a net increase in the continuum flux (a factor of
$\sim 1.17$ between the higher S/N 2002 FUSE observation and the 2010 COS observation)
and in the broad emission line flux between the two epochs, the overall shape of the
absorption troughs and continuum remains unchanged. Careful examination of the normalized \ovi\ COS and FUSE
absorption line profiles reveals troughs that are consistent, given the limited S/N of the FUSE observations, with no
variations between the epochs in any of the kinematic components. Therefore we use the \hi\ column densities
determined in the FUSE spectrum in our analysis. The \hi\ column density estimates reported in Table~\ref{coldensi} are
extracted from~\citet{Dunn10b} and consist of partial covering solutions derived on higher-order Ly-series lines.
We however note report the detection of the Ly10 line associated with kinematic component $T2$ revealing an
underestimation of the \hi\ column density by a factor of $\sim 2$ in \citet{Dunn10b}. 
The AOD solutions we report are computed on the Ly$\alpha$ line profile present in the COS spectrum and are
only lower limits given the saturation of the line profile. Finally, the absence of a bound-free edge for \hi\ in the
FUSE data allows us to place an upper limit on the \hi\ column density of $10^{16.2}$ cm$^{-2}$.

\subsubsection{\rm{\civ}}

Absorption troughs associated with \civ\ are found in all components of the outflow. The velocity
range of the outflow ($\sim 700~\mathrm{km}~ \mathrm{s}^{-1}$) being greater than the separation
between the components of the doublet \civ\ $\lambda \lambda$ 1548.200,1550.770 ($\sim 500~\mathrm{km}~ \mathrm{s}^{-1}$) limits the
possibility of deriving a partial covering and power law solution in several components of
the outflow because of self-blending between the red and blue lines of the doublet.
While the blue component of trough $T1$ is free of known blending, its red component is
blended by the blue component of trough $T4$. However, we note that the non-blended part of trough
$T4$ exhibit the 1:2 strength ratio between the doublet components as expected in the AOD model.
Assuming that the covering in trough $T4$ is not a strong function of velocity, blending
by the red $T1$ line is limited, suggesting that the trough $T1$ is also close to AOD.
Both red sub-components of trough $T2$ are severely blended by the blue
components of troughs $T5A$ and $T5B$. A lower limit on the column density can be placed on trough
$T2$ by computing the AOD solution on the non-blended blue line, while a lower limit on components
$T5A$ and $T5B$ is placed by the AOD solution on the non-blended red line. Kinematic component $T3$
is the only component where neither \civ\ of the lines are affected by self-blending allowing us to determine
the ionic column density using all three absorber models.

\subsubsection{\rm{\nv}}

Absorption troughs associated with \nv\ are observed in every kinematic component of the outflow.
Excepting the lower velocity section of trough $T4$, a high covering fraction indicated by the similarities in the column density
computed with the AOD and PC methods. The higher discrepancy observed for component $T1$ comes
from the fact that the blue \nv\ residual intensity is significantly below the expected level
assuming the AOD scenario. Since no blend is known to affect the blue \nv, this difference is
probably due to a slight overestimation of the emission model in that region, so that the 
column density determined on the red \nv\ line is probably a reliable measurement of the 
ionic column density in this component.

\subsubsection{\rm{\ovi}}

The \ovi\ troughs are located at the edge of the COS detector, where the
poor wavelength solution and lower S/N limit the constraints we can put on the
ionic column density. While a higher order correction of the wavelength solution
is probably needed at the edge of the detector, we use here a single velocity
shift of both the red and blue \ovi\ lines (respectively $12$ and $17 ~\mathrm{km}~ \mathrm{s}^{-1}$)
in order to align the core of the strongest kinematic components with the centroids
of the \nv\ $\lambda$ 1238.820 ones. This first order correction seems to be
sufficient for several components, however the match is not totally convincing
for others where a clear shift between the centroid of the blue and red line
persists (see Figure~\ref{alllineprof}).

Several components of both red and blue \ovi\ lines are blended by known ISM lines,
further limiting the accuracy of the column density estimates derived for this ion
in several components. The blue \ovi\ line is blended in component $T1$ by a weak
\feii\ $\lambda$1133.880 and by \ni\ $\lambda$1134.420 and $\lambda$1134.170 in trough $T2$
and by \ni\ $\lambda$1134.980 in trough T3. The troughs of stronger ISM lines associated with
\feii\ and \ni\ at longer wavelengths are shallow, indicating the blends only marginally
affect the line profiles of \ovi. In trough $T2$, the red 
line of \ovi\ is affected by an unidentified blend.

The integrated column densities derived for component $T1$ using the AOD and PC model
reveal a 60\% departure from AOD suggesting the PC gives a more realistic
estimate of the column. The same behavior is observed in component $T2$. 
The apparent  optical depth ratio between the
red and blue lines in components $T3$ and $T4$ is close to the expected laboratory
value, suggesting, like in the case of \nv, that the AOD determinations can be used.
Trough $T5B$ is totally saturated in its core and significantly saturated
in the wings ($I_r \sim I_b$). The AOD determination in this case is a strict lower limit
on the column density since even parts of the red line profile present a residual intensity
close to zero ($\tau \gg 1 $). The finite ionic column reported in Table~\ref{coldensi}
for component $T5B$ is derived using a maximum optical depth of $\tau = 4$ for these velocity bins.
Finally trough $T5A$ shows some evidence for partial covering effects. However, the presence
of an emission hump around $0 ~\mathrm{km}~ \mathrm{s}^{-1}$ suggests a possible underestimation
of the emission model at low velocities around the blue \ovi\ line, decreasing the
apparent departure from the AOD scenario for this component. In order to account for this
effect we use the AOD measurement on the red line as a lower limit on the column
density in component $T5A$.

\begin{figure}
  \includegraphics[angle=90,width=1.0\textwidth]{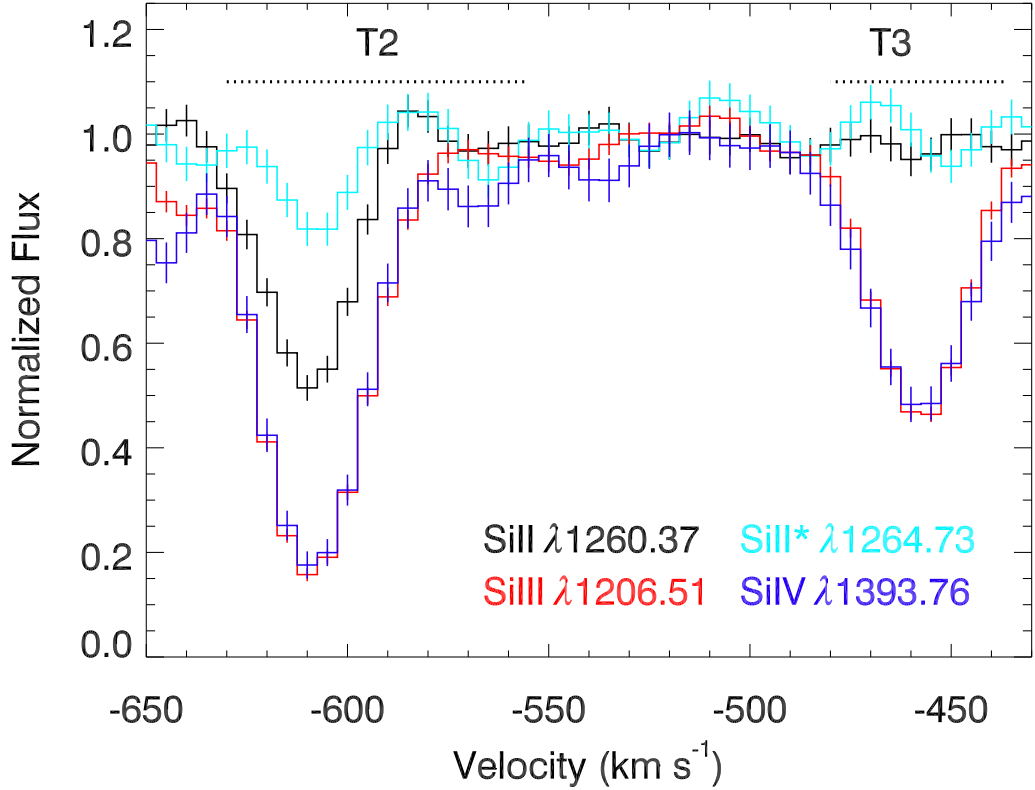}\\
 \caption{The observation of absorption troughs associated with three different ionization stages of the
Silicon atom allows us to better constrain the photo-ionization analysis of the absorber. We also note
that the residual intensity in the core of the \siIII\ $T2$ troughs match the saturated blue \siIV\ suggesting a
non-black saturation of the \siIII\ (see text). This behavior is also observed in trough $T3$, but leading to
a different conclusion due to the non saturation of \siIV\ in that case (see text).}
 \label{si_compa} 
\end{figure}

\subsubsection{\rm{\siIII}}
\label{silicon3}

\siIII\ signatures are identified in four of the kinematic components ($T2$, $T3$, $T5A$ and $T5B$) of the UV outflow.
A weak feature associated with component $T1$ may also be detected in the continuum noise (at less than the 2 $\sigma$ level), though, due to the limits of deconvolution, the deepness of the feature
is close to other ripples observed in the continuum and is probably a false detection. Given the nature of the detection,
we report an upper limit on the ionic column of \siIII\ for this kinematic component. For the other kinematic components,
we are only able to place a lower limit on the \siIII\ column given the impossibility of deconvolving the optical depth
from the covering fraction for singlet lines. We however note that the \siIII\ trough
associated with the kinematic component $T2$ shares a residual intensity identical to the one observed \siiv\ blue
(see Figures~\ref{alllineprof} and \ref{si_compa}). Given the non-black saturation noted in Section~\ref{silicon4} in component $T2$ for the
\siiv\ line, this observation suggests a net saturation of the \siIII\ trough whose deepness is mainly reflecting a partial
covering effect. We note that the residual intensity of the non saturated blue \nv\ line is similar to the one observed
in the saturated blue \siiv (see Figure~\ref{alllineprof}), which can be a coincidence or point to the fact that the PC
model does not constitute an appropriate model of the absorbing material distribution. The residual flux in component $T3$ of \siIII\ also shares
a similar depth with the blue \siiv\ line, however in this case the high covering fraction deduced from the residual fluxes observed in the \siiv\ doublet
lines prevent us from drawing a similar conclusion.

\subsubsection{\rm{\siiv}}
\label{silicon4}

\siiv\ troughs are identified in components $T2$, $T3$, $T5A$ and $T5B$. Trough $T2$ shows a significant departure 
from the AOD model suggesting a better description by an inhomogeneous model. Such effect is even stronger in component $T5A$ where
the red and blue \siiv\ line profiles matches almost perfectly over the whole component, only allowing us to 
derive a conservative lower limit on the column by assuming an optical depth limit of $\tau = 4$ in the saturated part
of the system. Trough $T3$ shows a high covering as revealed by the small difference between the ionic column densities
derived by the AOD and PC method. A similar behavior is also observed in trough $T5B$ but with a higher 
discrepancy in the columns due to the difficulty of getting reliable PC measurements for shallow troughs.

\subsubsection{\rm{\siv}}

 \siv\ is observed in kinematic component $T2$ as a shallow trough. The ionic column density is estimated
by applying the AOD method. A shallow feature (less than the 1 $\sigma$ level) in the normalized continuum
coincides with the expected position of the \siv\ line in trough $T3$ (see Figure~\ref{alllineprof}), but the feature is similar in depth to
other patterns observed in regions free of lines. For this reason we report an upper limit on the \siv\ line in trough $T3$
by fitting the template of the blue \siiv\ line to the $1 \sigma$ noise in that region.

\subsubsection{The density diagnostic lines}
\label{diagli}

  Absorption troughs associated with excited states of \siII\ and \cii\ are observed
in kinematic components $T2$ (both \siII\ and \cii) and $T3$ (only \cii) allowing us to determine the number density and hence
the distance to the outflowing material at their origin (see Section~\ref{distancesect}).

Four absorption troughs from the \siII\ resonance line ($\lambda$ 1190.42, 1193.28, 1260.37, 1526.72) free of obvious
contamination are identified within the COS range associated with the kinematic component $T2$ of the outflow. The weaker
$\lambda$ 1304.37 transition (detected at less than the 2 $\sigma$ level) is located in a region of lower S/N\footnote{Given the redshift of the IRAS F22456-5125,
the SiII $\lambda$ 1304.37 transition is located in a spectral region at the red edge of the G130M grating range and
at the blue edge of the G160M grating range.}, and is barely detected at the S/N level in that region.
Observation of four lines emanating from the same state and spanning a range of oscillator strengths allows us to further
investigate the absorber model by over-constraining the residual
intensity equations. In this case, we have four residual intensities to be fit by
two parameters (in the PC and PL models). However, evaluation of the fits to the data by these models requires the knowledge of reliable oscillator
strengths of each line.

\begin{figure}
  \includegraphics[angle=90,width=1.0\textwidth]{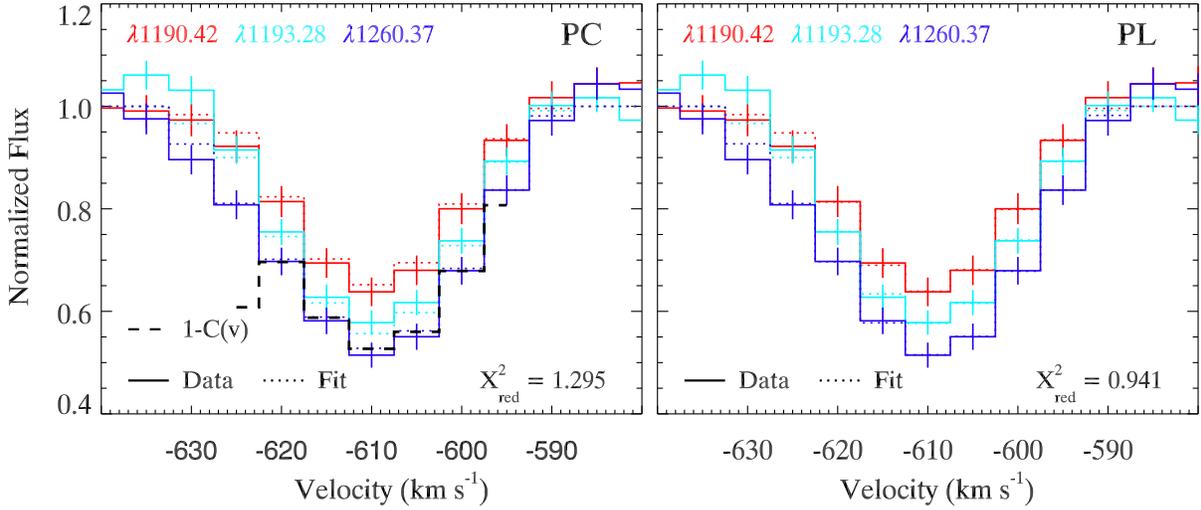}\\
 \caption{Partial Covering (left) and power law (right) model fit of three \siII\ resonance lines associated
with the kinematic component $T2$ of the UV outflow of IRAS F22456-5125. The original data and their error
are plotted in solid while the fitted fluxes are plotted in dotted line. The reduced $\chi^2$ value (see~\citealt{Arav08})
of the simultaneous fit for the three transitions is given in the bottom right of each panel. The Covering solution $C(v)$
is only presented in regions where the residual flux in the weakest lines does not affect the determination.}
 \label{siiireso} 
\end{figure}

Despite a number of theoretical studies, large uncertainties remain in the computed oscillator
strengths of the \siII\ transitions (see~\citealt{Bautista09} for details).
Using the oscillator strengths from NIST for the quoted transitions (rated either B+ or C in the database),
we find that the relative strength order of the lines matches the observed residual flux
for the $\lambda$ 1190.42, 1193.28 and 1260.37 lines and the weak detection of the $\lambda$ 1304.37 transition.
This is not the case for the $\lambda$ 1526.72 line, which is expected to be weaker than the $\lambda$ 1190.42
and $\lambda$ 1193.28 lines (hence having a larger residual flux), but for which we
observe a smaller residual intensity across the trough. The problem persists when using
the updated oscillator strengths reported in~\citet{Bautista09}. While this could be
due to a blend, the narrowness of the trough and its location away from any known ISM lines does not support
this scenario. For this reason, we use only the $\lambda$ 1190.42, 1193.28 and 1260.37 resonance lines to
compute the column density. We present in Figure~\ref{siiireso} results of the simultaneous fits of the three \siII\ lines
performed using the PC and PL absorber models. The PC model reveals a small covering
factor across the trough, ranging from 0.4 in the wings to 0.5 in the core. A better fit
to the observed line profiles is provided using the power law model. The derived power law exponent is close to $a=10$ across
the trough, corresponding to a peaked distribution suggesting that more than half of the
emission source is actually covered by optically thin material with $\tau < 0.1$. The results hold if we introduce the weak
 $\lambda$ 1304.37 transition in the computation, leading to changes in column densities that are less than 10\%
for both PC and PL absorber model. The column density derived using the PL absorber model is 2.5 times
larger than the one assuming the PC model, potentially suggesting an underestimation of the column density when using the PC model.
However, the PL method can overestimate the column density since a good
fraction of it originates in the optically thick region ($\tau > 4$) of the distribution, to which the residual flux is not
sensitive (see discussion in~\citealt{Arav05,Arav08}). Finally we note that the integrated column density over $T2$ only differs by
less than 7 \% when one uses the oscillator strength from NIST or the one reported in~\citet{Bautista09}.

 The strongest transition ($\lambda$ 1264.73) associated with the excited \siII\ (E=$287~ \mathrm{cm}^{-1}$, \siII* hereafter) is firmly identified
in kinematic component $T2$ (see Figure~\ref{si_compa}). A shallower absorption feature corresponding to the weaker excited transition ($\lambda$ 1194.50)
is distinguished at the S/N of the COS observations. Detection of two transitions from an ion with the same low energy level allows
us, in principle, to derive a velocity dependent solution of the column density using the PC or PL method. However the shallowness of
the troughs coupled to the limited signal to noise prevent us from computing a reliable solution across the trough. Nevertheless
we can still derive the column density associated with \siII* assuming a PC or PL model by using the velocity dependent solution of the covering
factor $C(v)$ or the power law parameter $a(v)$ derived above from the fitting of the resonance level transitions of the same ion.
With either $C(v)$ or $a(v)$ fixed in the equations of the residual intensity, we constrain the set of 2 equations for the
observed residual intensities in the \siII* lines and are able to derive the velocity dependent column density solution for both absorber models.
In trough $T3$, we put an upper limit on $N($\siII$) < 0.66~ 10^{12}$ cm$^{-2}$ due to non-detection (less that the 1 $\sigma$ level)
of the stronger $\lambda1260.37$ and $\lambda1264.73$ lines in the COS spectrum.

\begin{figure}
  \includegraphics[angle=90,width=1.0\textwidth]{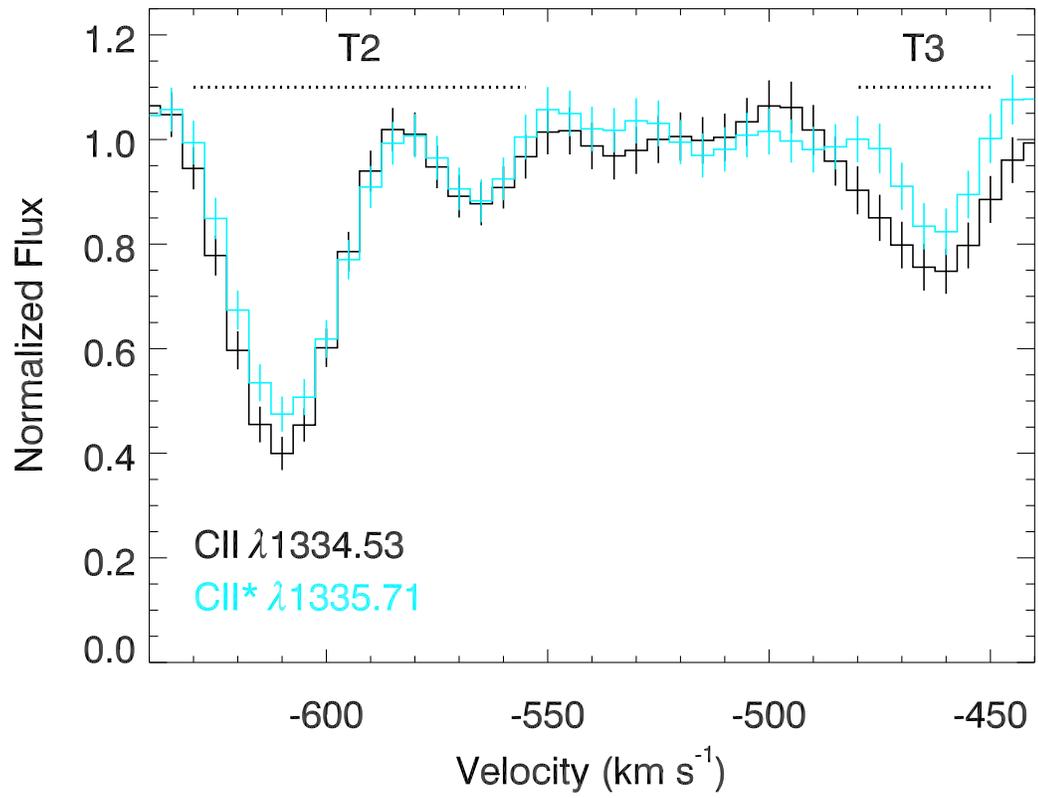}\\
 \caption{Line profile of \cii\ and \cii* rebinned to a common $5 ~\mathrm{km}~ \mathrm{s}^{-1}$ resolution velocity scale. Absorption troughs are identified in kinematic components $T2$ and $T3$ of the outflow.}
 \label{ciireso} 
\end{figure}

 \cii\ $\lambda$ 1334.53 (E=0 $cm^{-1}$) and the blend of \cii\ $\lambda$ 1335.66,1335.71 (E=63 $cm^{-1}$, hereafter \cii*) are detected in components $T2$ and $T3$ of the outflow.
Having only one line for each lower level does not allow us to deconvolve the effects of partial covering and population ratio
of the level, allowing us to only provide an AOD estimate. In the case of \siII\ we saw that the covering derived for that ion
was quite small, having a covering close to 0.5 in the core of the trough. Looking at the residual intensity in the core of the \cii\ line profile in Figure~\ref{ciireso}, 
it is clear that the covering of that line is larger than 0.5 
Thus using the \siII\ covering solution does not allow us to reproduce the observed \cii\ profiles. In order to tentatively estimate the effect of covering
on the \cii\ and \cii* columns, we compute the ionic column density using the covering solution derived from \siiv, a medium ionization species, and report it in 
Table~\ref{coldensi}. While we observe a small increase of the derived columns using this PC model, the ratio of column density between the resonance and excited states
remains identical (as expected given the similar residual flux inside the \cii\ and \cii* troughs), strengthening the density diagnostic obtained
from these lines.

\section{Photoionization analysis of the absorbers}
\label{analiab}

In order to derive the physical properties of each kinematic component of the outflow, we
solve the ionization equilibrium equations using version c08.00 of the spectral synthesis
code \textsc{Cloudy} (last described by \citealt{Ferland98}). We model each absorber by a
plane-parallel slab of gas of constant hydrogen number density ($\vy{n}{H}$) and assume 
solar elemental abundances as given in \textsc{Cloudy}. The spectral energy distribution 
(SED) we use was described in \citet{Dunn10b}. Using the grid-model approach described in 
Paper I, we find a combination of total hydrogen column density ($N_H$) and ionization parameter 
that best reproduces the observed ionic column densities reported in Table~\ref{coldensi}.
The ionization parameter depends on the distance ($R$) to the absorber from the central source and is given by
\begin{equation}
\label{eqioni}
U_H = \frac{Q_H}{4 \pi R^2 \vy{n}{H} c},
\end{equation}
where $Q_H =$ 2.5 x 10$^{55}$ s$^{-1}$ is the rate of hydrogen ionizing photons emitted by
the object, and $c$ is the speed of light. We estimate the hydrogen ionizing rate $Q_H$ (and also the bolometric luminosity $L_{Bol}$)
by matching the flux of the model SED to the de-reddened observed flux at 1150 \AA (rest-frame) using a standard cosmology
($H_0$=73.0 km s$^{-1}$ Mpc$^{-1}$, $\Omega_{\Lambda}$=0.73, $\Omega_m$=0.27).

The COS observations show a wealth of absorption lines compared to the earlier
FUSE observations discussed in~\citet{Dunn10b}. This allows us to derive
more accurate physical properties of the absorbing clouds associated with the
UV outflow. In the following subsections, we describe the photoionization solution derived
for each kinematic component.  As discussed in Section~\ref{hicold}, the physical properties of the absorber do not
appear to change between the FUSE and COS epochs. Therefore, we use the column densities
of \hi\ and \ciii\ derived in \citet{Dunn10b} from FUSE data.

We characterize the maximum likelihood for the model of each kinematic component by the
merit function :
\begin{equation}
  \chi^2 =  \displaystyle\sum\limits_{i}  \left(  \frac{   \log{N_{i,mod}}-\log{N_{i,obs}}   }{ \log{ N_{i,obs} } - \log{ ( N_{i,obs} \pm \sigma_i )}    }   \right)^2,       
\end{equation}
where, for ion $i$, $N_{i,obs}$ and $N_{i,mod}$ are the observed and modeled column densities, respectively, and $\sigma_i$ is the error in the measured column density.
We prefer this formalism to the traditional definition of $\chi^2$ since it preserves the multiplicative nature of the errors when dealing with logarithmic values.

\subsection{Troughs $T2$ and $T3$}
\label{pisolt3}

\begin{figure}
  \includegraphics[angle=0,width=1.0\textwidth]{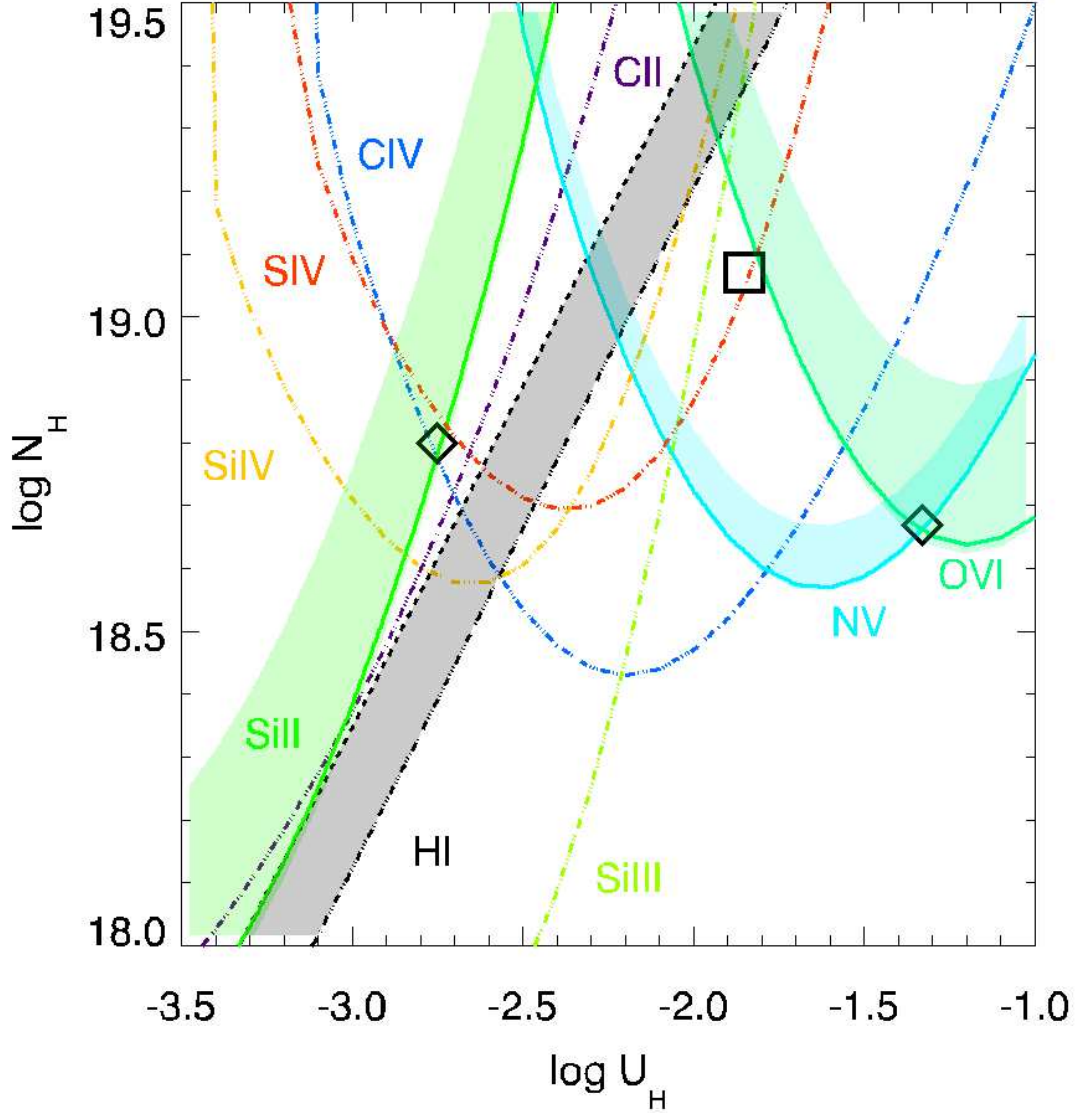}\\
 \caption{Photoionization modeling of kinematic component $T2$. The plotted lines represent
slab models whose predicted $N_{ion}$ matches the estimated values. Solid lines indicate a
measured column density, dotted dashed lines a lower limit on the column density and dotted lines
an upper limit on the column density. The error bars due to the photon noise as well as the
systematic uncertainties in the absorber model are represented as a shaded area for each ions when an estimate is
available. The black diamonds marks the two ionization component model that best fits the estimated $N_{ion}$ while
the black square marks our best single ionization model.}
 \label{T2PI} 
\end{figure}

The physical parameters of component $T2$ are constrained by ten ionic column densities, 
eight from COS data along with \hi\ and \ciii\ from FUSE data (keeping in mind that the latter
have been obtained at a different epoch). The ions span a wide range of ionization stages 
from \cii\ and \siII\ to \nv\ and \ovi. A plot of the
results for a grid of photoionization models for trough T2 is presented in Figure~\ref{T2PI}.
The least-squares single-ionization parameter solution ($\chi^2 = 1147$) is marked with a square in the $N_H-U_H$
plane, and predicted values for ionic column densities are given in Table~\ref{t2__cloudy_model}. 
The \cii\ and \siII\ column densities predicted by that model are underestimating the observed
column densities by one and two order of magnitude, respectively, therefore making the model physically unacceptable. Due to the poor fit
of this model to the data, we use a two-ionization component model ($\chi^2 = 6.7$), which is depicted by diamonds in Figure~\ref{T2PI}.
All of  the ions from the COS data are fit well with the two-component model (see Table~\ref{t2__cloudy_model}).


\begin{deluxetable}{lccccc}
 \tablewidth{0.8\textwidth}
 \tablecolumns{6}
 \tabletypesize{\footnotesize}
 \tablecaption{Photoionization models for component $T2$}
 \tablehead{
 \colhead{Ion} &
 \colhead{log($N_{ion}$) (cm$^{-2}$)} &
\colhead{log$\left(\frac{N_{mod}}{N_{obs}}\right)$} &
\multicolumn{2}{c}{log $N_{mod}$} &
\colhead{log$\left(\frac{N_{mod}}{N_{obs}}\right)$} \\
 \colhead{} &
 \colhead{Adopted $^{\mathrm{a}}$} &
 \colhead{SI$^{\mathrm{b}}$} &
 \colhead{TI$_{lo}^{\mathrm{b}}$} &
 \colhead{TI$_{hi}^{\mathrm{b}}$} &
 \colhead{TI$_{lo}$+TI$_{hi}$}
  }
 \startdata
 log $U_H$ & $\cdots$ & -1.9 & -2.8 & -1.3 & \\
 log $N_H$ & $\cdots$ & 19.1 & 18.8 & 18.7 &  19.1\\
\hline

 \hi\   & $\in [$ 15.97, 16.19 $]$ & -0.25 & 16.38 & 14.66 & +0.20\\
 \cii\  & $\gtsim$ 13.95                & -1.16 & 14.13 & 10.76 & +0.18\\
 \civ\  & $>$ 14.40                & +0.54 & 14.32 & 13.96 & +0.08\\
 \nv\   & 14.07$^{+0.10}_{-0.01}$  & +0.40 & 12.80 & 14.05 &  0.00\\
 \ovi   & 14.91$^{+0.25}_{-0.02}$  & -0.18 & 12.18 & 14.91 &  0.00\\
\siII\  & 13.17$^{+0.37}_{-0.03}$  & -1.91 & 13.23 & 8.45  & +0.06\\
\siIII\ & $>$ 12.96                & -0.18 & 14.01 & 10.25 & +1.05\\
\siIV\  & $>$ 13.67                & -0.37 & 13.85 & 11.14 & +0.18\\
\siv\   & $\gtsim$ 13.69           & +0.11 & 13.59 & 12.05 & -0.09\\

 \enddata
 \label{t2__cloudy_model}
 a) Adopted column densities reported in Table~\ref{coldensi}.
 b) The label SI corresponds to the single ionization model while TI$_{lo}$ and  TI$_{hi}$
    are the low and high ionization phase of the two ionization model of the absorber.
\end{deluxetable}

\begin{deluxetable}{lccccc}
 \tablewidth{0.8\textwidth}
 \tablecolumns{6}
 \tabletypesize{\footnotesize}
 \tablecaption{Photoionization models for component $T3$}
 \tablehead{
 \colhead{Ion} &
 \colhead{log($N_{ion}$) (cm$^{-2}$)} &
\colhead{log$\left(\frac{N_{mod}}{N_{obs}}\right)$} &
\multicolumn{2}{c}{log $N_{mod}$} &
\colhead{log$\left(\frac{N_{mod}}{N_{obs}}\right)$} \\
 \colhead{} &
 \colhead{Adopted $^{\mathrm{a}}$} &
 \colhead{SI$^{\mathrm{b}}$} &
 \colhead{TI$_{lo}^{\mathrm{b}}$} &
 \colhead{TI$_{hi}^{\mathrm{b}}$} &
 \colhead{TI$_{lo}$+TI$_{hi}$}
  }
 \startdata
 log $U_H$ & $\cdots$ & -2.0 & -2.7 & -1.7 & \\
 log $N_H$ & $\cdots$ & 19.0 & 18.2 & 18.8 & 18.90\\
\hline

 \hi\   & 15.63$^{+0.08}_{-0.08}$  & +0.10 & 15.69 & 15.20 & +0.19\\
 \cii\  & $>$ 13.39                & -0.45 & 13.42 & 11.97 & +0.05\\
 \civ\  & 14.53$^{+0.17}_{-0.01}$  & +0.36 & 13.85 & 14.49 & +0.05\\
 \nv\   & 14.22$^{+0.27}_{-0.02}$  & +0.08 & 12.44 & 14.25 & +0.04\\
 \ovi   & 14.81$^{+0.01}_{-0.01}$  & -0.35 & 11.93 & 14.72 & -0.09\\
\siII\  & $<$11.80$^{\mathrm{c}}$  & -0.29 & 12.47 & 10.23 & +0.67\\
\siIII\ & $>$ 12.58                & +0.38 & 13.36 & 11.84 & +0.79\\
\siIV\  & 13.67$^{+0.07}_{-0.03}$  & +0.21 & 13.27 & 12.49 & +0.13\\
\siv\   & $<$ 13.38                & +0.44 & 13.06 & 13.18 & -0.04\\

 \enddata
 \label{t3__cloudy_model}
 a) Adopted column densities reported in Table~\ref{coldensi}.
 b) The label SI corresponds to the single ionization model while TI$_{lo}$ and  TI$_{hi}$
    are the low and high ionization phase of the two ionization model of the absorber.
 c) Upper limit set by the non detection of the stronger \siII\ and \siII* in that component.
\end{deluxetable}

\begin{figure}
  \includegraphics[angle=0,width=1.0\textwidth]{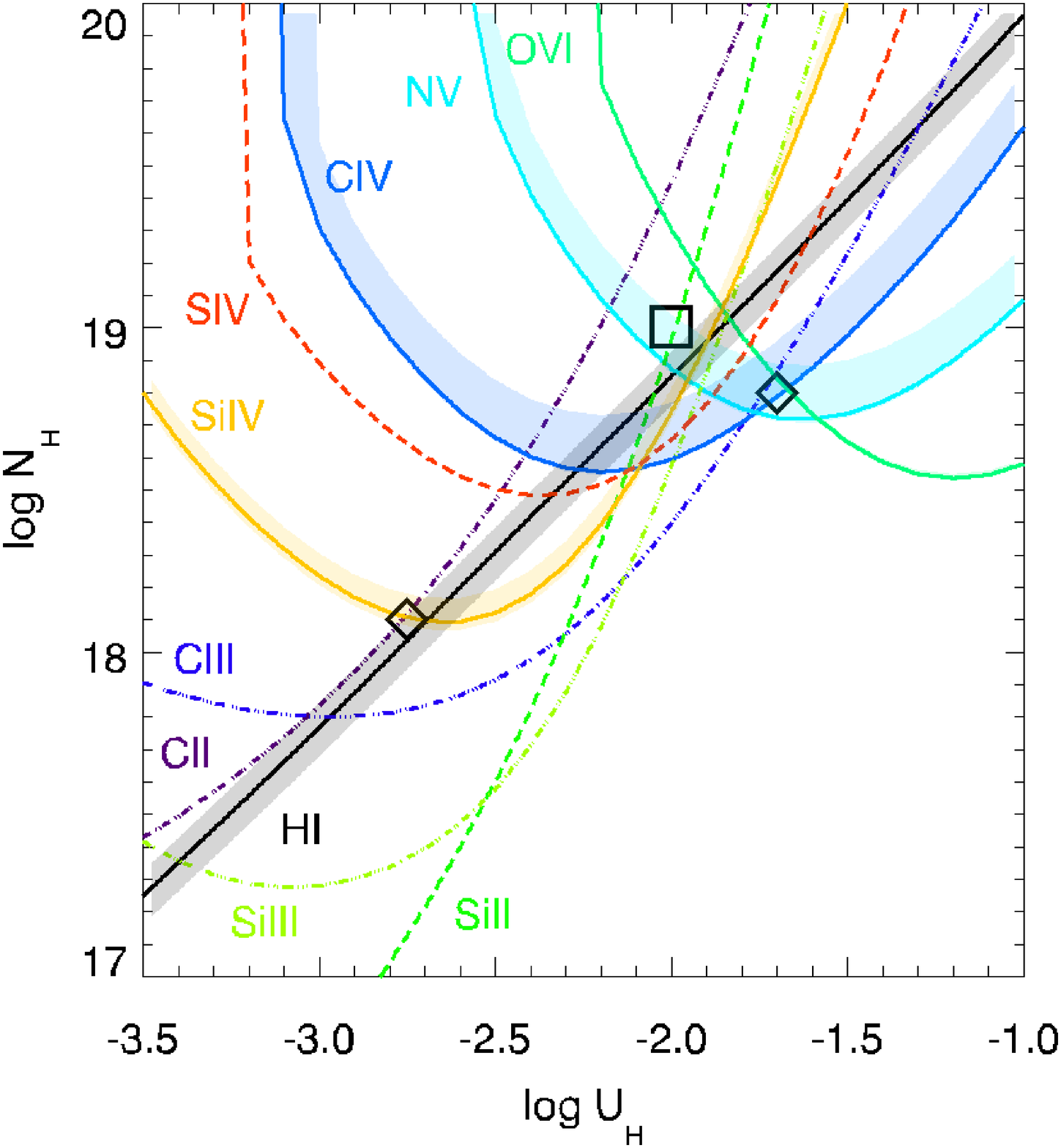}\\
 \caption{Photoionization solutions to trough $T3$. Similar presentation as Figure~\ref{T2PI}.}
 \label{T3PI} 
\end{figure}

For component $T3$, we have column density measurements for seven ions in the COS spectrum, along
with \hi\ and \ciii\ from FUSE data and an upper limit on \siII\ due to non-detection 
of the stronger lines in the COS spectrum (see Section~\ref{diagli}). A grid-model for trough $T3$ is plotted in Figure~\ref{T3PI}.
A single-ionization parameter solution ($\chi^2 = 1008$) is marked with a square in the $N_H-U_H$ plane. This solution 
fits all the lines within a factor of $\sim 3$ (see Table~\ref{t3__cloudy_model}). An improvement to the fit ($\chi^2 = 5.21$) for most ions is provided 
by the two-ionization parameter solution marked with diamonds in Figure~\ref{T3PI}, with the exception 
being an over-prediction of \siII\ by a factor $\gtsim 4$ (see Table~\ref{t3__cloudy_model}).

\begin{figure}
  \includegraphics[angle=0,width=1.0\textwidth]{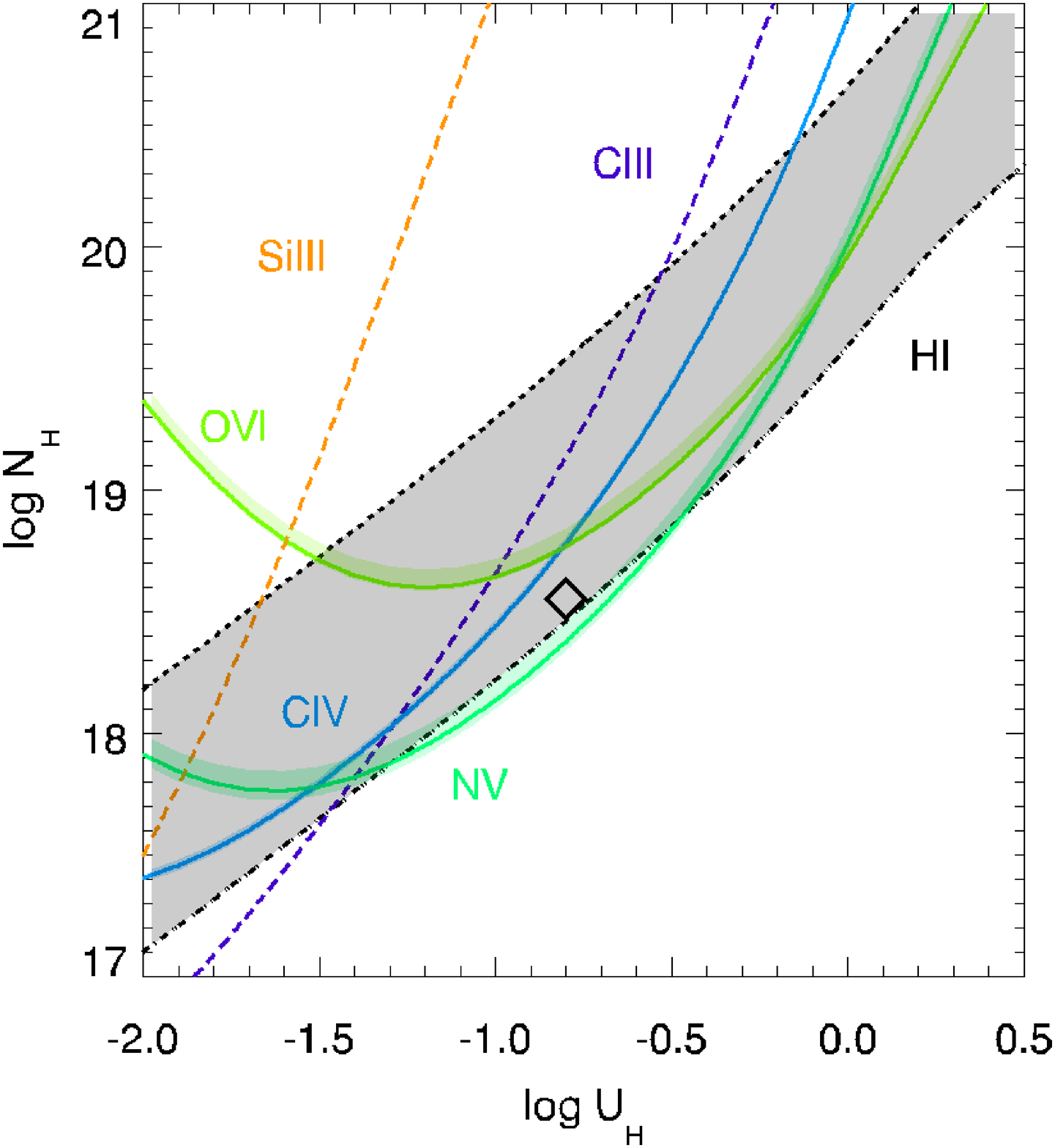}\\
 \caption{Photoionization modeling of kinematic component $T1$. Identical description to Figure~\ref{T2PI}.}
 \label{T1PI} 
\end{figure}

\begin{figure}
  \includegraphics[angle=0,width=1.0\textwidth]{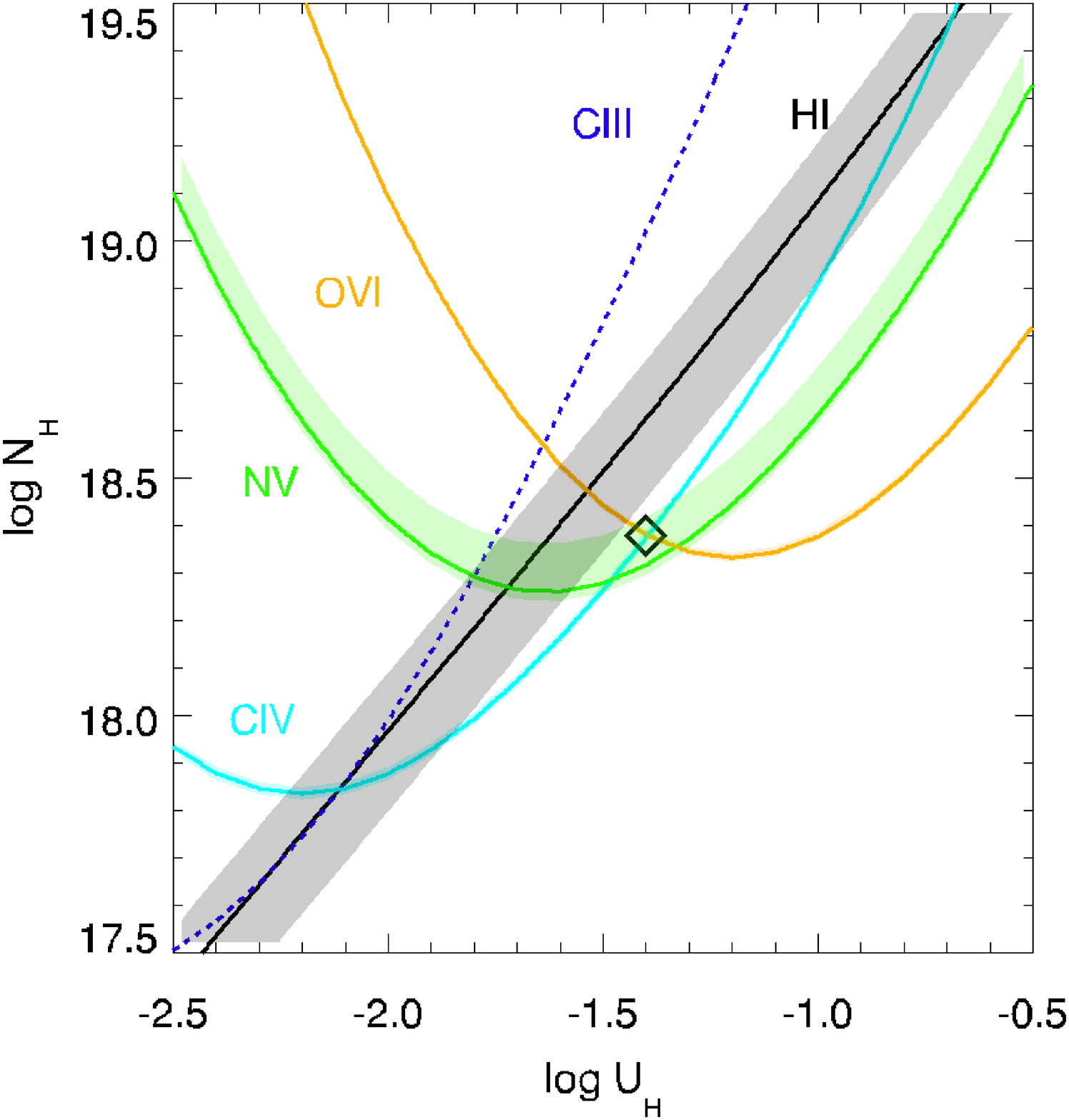}\\
 \caption{Photoionization modeling of kinematic component $T4$. Identical description to Figure~\ref{T2PI}.}
 \label{T4PI} 
\end{figure}

\begin{figure}
  \includegraphics[angle=0,width=1.0\textwidth]{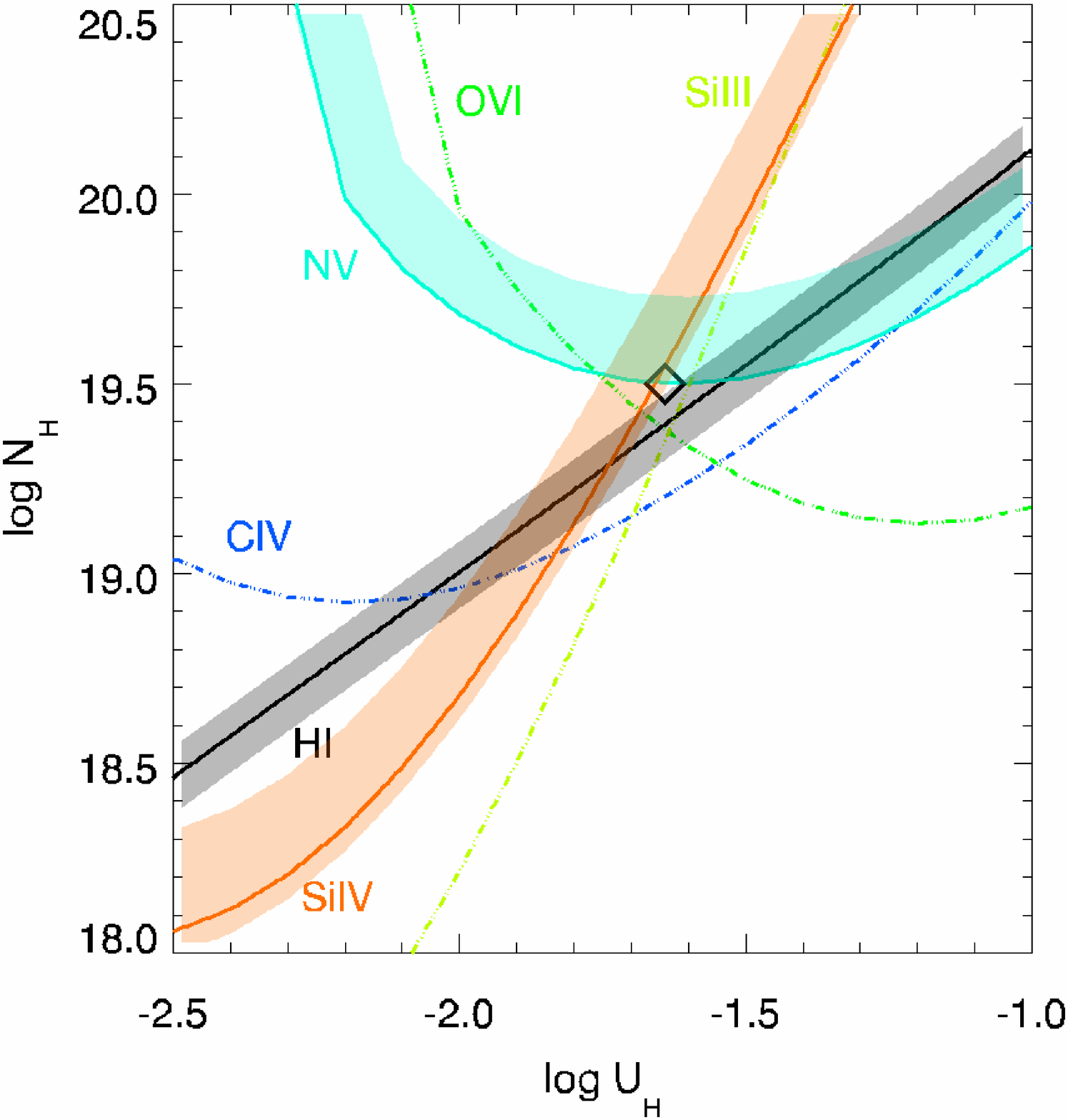}\\
 \caption{Photoionization modeling of kinematic component $T5B$. Identical description to Figure~\ref{T2PI}.}
 \label{T5BPI} 
\end{figure}

\begin{figure}
  \includegraphics[angle=0,width=1.0\textwidth]{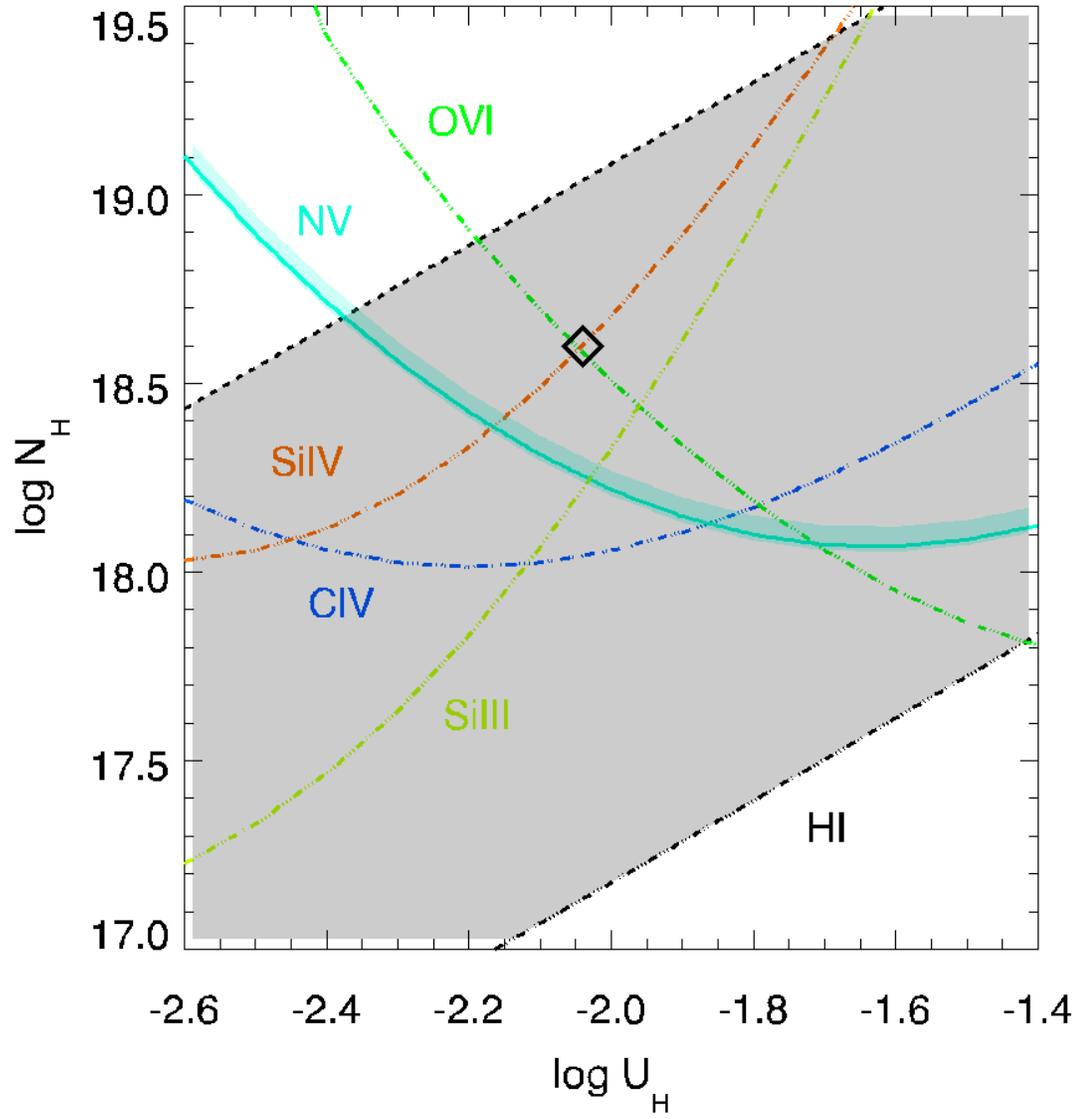}\\
 \caption{Photoionization modeling of kinematic component $T5A$. Identical description to Figure~\ref{T2PI}.}
 \label{T5API} 
\end{figure}

\subsection{Troughs $T1$, $T4$, $T5A$ and $T5B$}

For kinematic component $T1$, we essentially have five constraints (\civ, \nv, \ovi, \siIII\ and \hi) defining the region
of the ($N_H$,$U_H$) parameter space able to reproduce the estimated ionic column densities.
Visual inspection of Figure~\ref{T1PI} suggests a solution around $\log U_H \sim -0.8$
and $\log N_H \sim 18.6$, consistent with the upper limits on \ciii\ from~\citet{Dunn10b}.
This solution ($\chi^2 = 86$) accounts for every constraint to within 0.25 dex. A better solution can be found by relaxing the constraint of solar metallicity.
Considering the scaling of elemental abundances of C, N and O as a function of the metallicity $Z$~\citep{Hamann93,Hamann97}, 
an improved solution ($\chi^2 = 0.5$) is found for a gas of sub-solar metallicity ($[Z/Z_{\sun}] \sim$ -0.4) using
an identical $\log U_H$ and total hydrogen column of $\log N_H \sim 19.1$ for the slab.

\begin{deluxetable}{lcccc}
 \tablewidth{0.8\textwidth}
 \tablecolumns{5}
 \tabletypesize{\footnotesize}
 \tablecaption{{\sc Ionization solution for each kinematic component}}
 \tablehead{
 \colhead{Component} &
 \multicolumn{2}{c}{log($U_H$)$^{\mathrm{a}}$} &
 \multicolumn{2}{c}{log($N_H$)$^{\mathrm{a}}$ ($\mathrm{cm}^{-2}$)} \\
 \colhead{} &
\colhead{This work}&
\colhead{Dunn10$^{\mathrm{b}}$}&
\colhead{This work}&
\colhead{Dunn10$^{\mathrm{b}}$}
  }
\startdata

$T1$                            &  -0.8$^{+0.24}_{-0.20}$  & -1.15 &  18.6$^{+0.43}_{-0.21}$     & 18.6 \\
$T2$ SI$^{\mathrm{c}}$          &  -1.9$^{+0.16}_{-0.15}$  & -1.30 &  19.1$^{+0.26}_{-0.24}$     & 19.7 \\
$T2$ TI$_{lo}^{\mathrm{d}}$     &  -2.8$^{+0.04}_{-0.05}$  &  ...  &  18.8$^{+0.03}_{-0.04}$     & ...  \\
$T2$ TI$_{hi}^{\mathrm{d}}$     &  -1.3$^{+0.10}_{-0.09}$  &  ...  &  18.7$^{+0.07}_{-0.04}$     & ...  \\
$T3$ SI$^{\mathrm{c}}$          &  -2.0$^{+0.15}_{-0.14}$  & -1.27 &  19.0$^{+0.18}_{-0.18}$     & 19.7 \\
$T3$ TI$_{lo}^{\mathrm{d}}$     &  -2.7$^{+0.08}_{-0.06}$  &  ...  &  18.2$^{+0.07}_{-0.07}$     & ...  \\
$T3$ TI$_{hi}^{\mathrm{d}}$     &  -1.7$^{+0.03}_{-0.02}$  &   ... &  18.8$^{+0.02}_{-0.03}$     &  ... \\
$T4$                            &  -1.4$^{+0.01}_{-0.01}$  & -1.54 &  18.4$^{+0.01}_{-0.01}$     & 18.5 \\
$T5B$                           &  -1.6$^{+0.04}_{-0.04}$  & -1.05 &  19.5$^{+0.06}_{-0.01}$     & 20.1 \\
$T5A$                           &  -2.0$^{+0.04}_{-0.04}$  &  ...  &  18.6$^{+0.09}_{-0.10}$     &  ... \\

\enddata
\label{pi_solu_table}
 a) The error we report on our determinations of $U_H$ and $N_H$ are estimated by size of the contour in the ($U_H$, $N_H$) plane of the solutions
that have a $\chi^2$ value twice the $\chi^2$ of the best fit solution.
 b) From \citet{Dunn10b}.
 c) The label SI corresponds to the single ionization model. 
 d) TI$_{lo}$ and  TI$_{hi}$ are the low and high ionization phase of
    the two ionization model of the absorber.

\end{deluxetable}

The constraints on the ($N_H$,$U_H$) parameter space for trough $T4$ are presented in Figure~\ref{T4PI}. A solution
consistent with the measured ionic column densities is found near $\log U_H \simeq -1.4$ and $\log N_H \simeq 18.4$ ($\chi^2 = 2.9$). The
solution is suggesting roughly solar abundances of the gas though the slight discrepancy between the measured
\hi\ column density and that predicted by the solution may suggest a sub-solar metallicity medium.

The photoionization solution derived for trough $T5B$ and $T5A$ are presented in Figures~\ref{T5BPI} and \ref{T5API},
respectively. Inspection of Figure~\ref{T5BPI} suggests a solution around $\log U_H \sim -1.6$ and $\log N_H \sim 19.5$ ($\chi^2 = 3.6$)
for component $T5B$. A least-square fit to the constraints available for trough $T5A$ (Figure~\ref{T5API}) provides
a solution near $\log U_H \simeq -2.0$ and $\log N_H \simeq 18.6$ ($\chi^2 = 27$). While the saturation observed in the troughs of
several ions limits the analysis of the physical properties of the gas, the estimated ($N_H$,$U_H$) solution
is able to reproduce most of the ionic columns to within a factor of 2.

\section{Absorber distance and energetics}
\label{distancesect}

Detection of resonance and excited state transitions from \siII\ and \cii\ in troughs $T2$ and $T3$ allows us
to determine the distance to these two kinematic components from the central source. As can be seen from the definition of the
ionization parameter $U_H$ (Equation~\ref{eqioni}),
knowledge of the hydrogen number density $n_H$ for a given $U_H$ and $Q_H$ allows us to derive the distance $R$. When
an excited state is populated by collisional excitation, the population of that state compared to the resonance
level depends on the electron number density $n_e$~\citep{AGN^3}, which is $\sim 1.2~ n_H$ in highly ionized
plasma. Note that photoexcitation could also populate the metastable levels, however, with an IR flux $\sim 0.1$~Jy, population of the metastable
levels of \cii\ and \siII\ is negligible in IRAS~F22456-5125.

\begin{figure}
  \includegraphics[angle=90,width=1.0\textwidth]{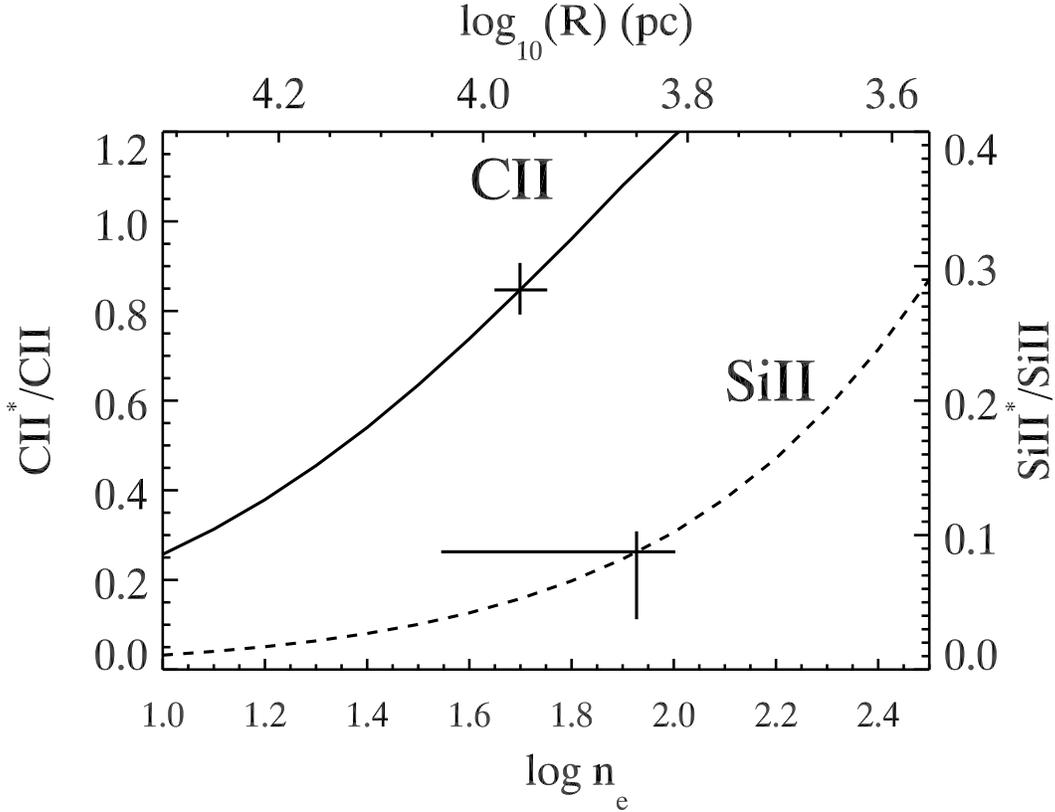}\\
 \caption{Density diagnostic for kinematic component $T2$. In this figure we plot the theoretical ratio of the level population
of the first excited states of \cii\ ($E=63~ \mathrm{cm}^{-1}$) and of \siII\ ($E=287~ \mathrm{cm}^{-1}$) to the level population
of the ground state versus the electron number density $n_e$ for a temperature of 10000 K (the diagnostic is relatively insensitive to temperature
for temperatures typical in UV absorber). The ratios derived from \cii\ and \siII\ (black crosses) imply an electron number density $\log n_e \simeq 1.7$. The uncertainty on the derived $n_e$
only accounts for the error on the ratio $N_{ion*}/N_{ion}$. On the top axis, we report the corresponding distance as a function of $n_e$
considering the ionization parameter of the low ionization phase ($\log U_H=-$2.8). 
}
 \label{t2_exci} 
\end{figure}

In trough $T2$ we observe resonance and excited states from \cii\ and \siII\ for which column densities have been derived in 
Section~\ref{diagli} and reported in Table~\ref{coldensi}. In Figure~\ref{t2_exci} we compare computed collisional excitation models
for \cii\ and \siII\ to the measured ratio of the column density between the excited and ground state of these two ions. For
the \siII*/\sii\ ratio, the large uncertainty comes from using PC and PL measurements of the column density
associated with \siII* and \sii. The \cii*/\cii\ ratio is consistent with the \siII*/\sii\ ratio. Given the similar but still
significantly different residual flux observed in the \cii* and \cii\ line profile, while the value of the column density
associated with \cii* and \cii\ can change with the different absorber model, their ratio is less affected since both
column density will scale with a similar factor (see Section~\ref{diagli}). For this reason we use the \cii*/\cii\
ratio to derive the electron number density of $\log n_e \simeq$ 1.70$^{+0.30}_{-0.15}$ for the low ionization phase of component $T2$. 
Using the derived ionization parameter of that phase, this density implies a distance
of $R \simeq 10.3^{+5.1}_{-1.6}$ kpc where the errors are conservatively computed from the $n_e$ range allowed
by the \siII*/\siII\ ratio and the error on the ionization parameter. We note that the density derived for this component is consistent with the non
detection of \siv*, only expected to be observed at higher densities.

 \begin{deluxetable}{lrrrrrr}
\tablecaption{{\sc Physical properties of the two kinematic components $T2$ and $T3$.}}
\tablewidth{0pt}
\tablehead{
\colhead{}
&\colhead{log $U_H$}
&\colhead{log $N_H$ }
&\colhead{$\log n_e$}
&\colhead{$R$ }
&\colhead{$\dot{M}_T$ }
&\colhead{$\log \dot{E}_k$}\\
\colhead{}
&\colhead{}
&\colhead{$\mathrm{cm}^{-2}$}
&\colhead{$\mathrm{cm}^{-3}$}
&\colhead{kpc}
&\colhead{M$\odot$/yr}
&\colhead{ergs/s}

}

\startdata

$T2$       &    -2.8$^{+0.04}_{-0.05}$      &    18.8$^{+0.03}_{-0.04}$  &  1.70$^{+0.30}_{-0.15}$  &  \multirow{2}{*}{ {\huge \}} 10.3$^{+5.1}_{-1.6}$}  & \multirow{2}{*}{5.1$^{+2.6}_{-0.9}$}  & \multirow{2}{*}{41.8$^{+0.2}_{-0.1}$} \\
           &    -1.3$^{+0.10}_{-0.09}$      &    18.7$^{+0.07}_{-0.04}$  &  0.20$^{+0.31, \mathrm{a}}_{-0.19}$  &                                        &                                        &                                        \\
$T3$       &    -2.7$^{+0.08}_{-0.06}$      &    18.2$^{+0.07}_{-0.07}$  &  1.20$^{+0.12}_{-0.10}$  &  \multirow{2}{*}{ {\huge \}} 16.3$^{+3.1}_{-1.9}$}  & \multirow{2}{*}{4.1$^{+0.9}_{-0.7}$}  & \multirow{2}{*}{41.4$^{+0.1}_{-0.1}$} \\
           &    -1.7$^{+0.03}_{-0.02}$      &    18.8$^{+0.02}_{-0.03}$  &  0.20$^{+0.14, \mathrm{a}}_{-0.12}$  &                                        &                                        &                                        \\

\enddata

\label{tabmflo}
a - Computed using the low ionization component number density and assuming that the high ionization component is located at the same distance
from the central source (see Section~\ref{discussion})\\

\end{deluxetable}

\begin{figure}
  \includegraphics[angle=90,width=1.0\textwidth]{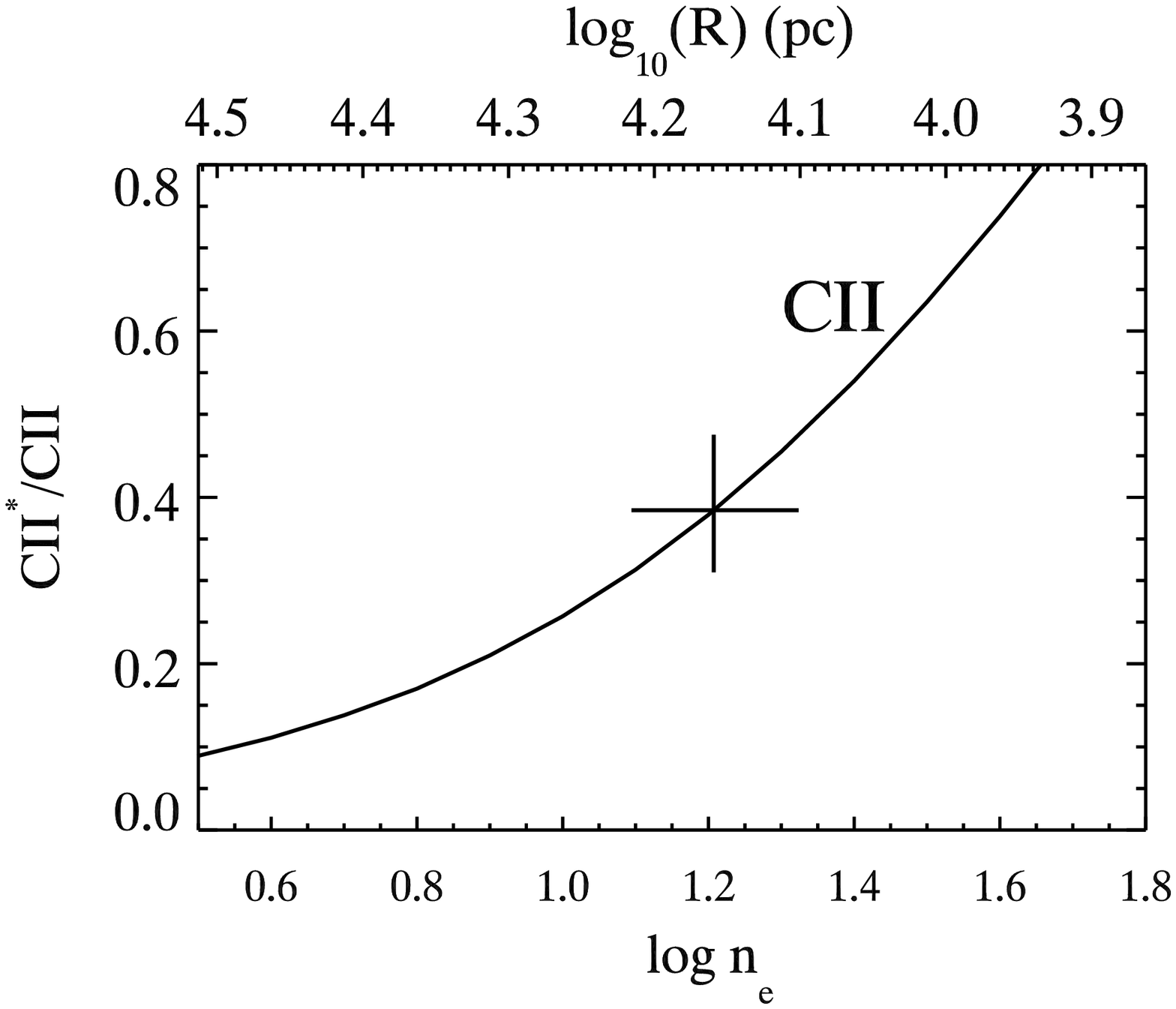}\\
 \caption{Density diagnostic for kinematic component $T3$. We plot the theoretical ratio of the level population
of the first excited states of \cii\ ($E=63~ \mathrm{cm}^{-1}$) to the level population
of the ground state versus the electron number density $n_e$ for a temperature of 10000 K. The ratio
derived from \cii\ and \cii* lines is implying an electron number density $\log n_e \simeq 1.2$. On the top axis, we report the corresponding distance as a function of $n_e$
considering the ionization parameter of the low ionization phase ($\log U_H=-$2.7).}
 \label{t3_exci} 
\end{figure}

In trough $T3$, the only excited state we observe is associated with the \cii* transitions. Comparing the computed collisional excitation models
for \cii\ to the measured ratio of the column density between the excited and ground states of this ion (see Figure~\ref{t3_exci}) we find
$\log  n_e \simeq$ 1.20$^{+0.12}_{-0.10}$ for the low ionization gas phase producing the \cii\ and \cii* troughs.  Given the photoionization
solution for that phase quoted in Section~\ref{pisolt3} the derived electron number density imply a distance of $R \simeq 16.3^{+3.1}_{-1.9}$ kpc
where once again the errors on $R$ reflect the uncertainty in $n_e$ and the ionization parameter. Using the single ionization solution for component $T3$ (see Table~\ref{pi_solu_table})
leads to a distance estimate reduced by a factor of $\sim$2 to $R \simeq 7.3^{+1.9}_{-1.3}$ kpc. Note that in the case of $T3$, we consider the AOD determination of the column
densities of \cii* and \cii\ given the absence of multiple lines allowing to test for the absorber model. Similarly to what is observed for the
\siII\ transitions in kinematic component $T2$, the net difference in residual flux between the excited and resonance troughs could lead an inhomogeneous absorber
model to predict a smaller ratio (a factor of two in the case of the \siII*/\siII\ ratio of trough $T2$). The \cii*/\cii\ ratio derived here
could then be viewed as an upper limit on the true ratio, the latter being possibly overestimated by up to a factor of $\sim 2$, leading to an 
underestimation of the distance $R$ by a factor of $\sim \sqrt{2}$.

The reported distances are very large compared to the size of the emission regions in AGNs, where we estimate the broad line
emission region to be roughly 0.03 pc in scale \citep{kaspi05}. Similar distances to narrow absorption line system
intrinsic to quasars and Seyferts exhibiting lines from excited states of low ionization species have already been
reported in the literature (e.g. \citealt{Hamann01}, \citealt{Hutsemeker04}, Paper I). The kinematic structure and
timescale deduced from the line profiles lends support to the interpretation of these narrow absorption line systems
as associated with episodes of mass ejection rather than a continuous wind currently (in the AGN rest-frame) emanating
from the central regions. In the simplest geometrical picture, we can naively estimate the thickness of
each shell of ejected material by $\Delta R = N_H/n_H$, where $n_H$ is the hydrogen number density of the low density, high ionization phase, giving a
$\Delta R < 2$ pc for both components $T2$ and $T3$. This is assuming that the individual kinematic components can be described as a uniform
slab having an internal volume filling factors $f$ of unity. This situation is nonphysical since for the inferred temperature of the
absorbing gas ($T \sim 10^4$ K) the velocity width of the outflow ($\Delta v \gtsim 50$ km s$^{-1}$)is at least ten times
larger than the sound speed, and therefore the outflowing material cannot be a sonically connected entity. The large $\Delta v$
of this highly supersonic outflow is then probably due to bulk motion of the absorbing material. Assuming that the outflowing material is not
decelerating, we can obtain the dynamical timescale of the shell $t_{s} = R/v_{s} \sim $ 20 Myr, where we choose an average
outflow speed of $v_s = 500$ km s$^{-1}$. Over these 20 Myr taken by the shell to reach the distance $R$, it has been expanding
at a speed $v_{exp} = FWHM \sim v /10$ (see Section~\ref{identi}), so that $\Delta R \sim 0.1 R$. We can use this thickness in
order to estimate the actual filling factor $f$ of the shell since $\Delta R = N_H/(f n_H)$. Using the $N_H$ and $n_H$ of the high ionization phase reported
for component $T2$ and $T3$ we find $f \sim 10^{-3}$. This number is in agreement with the low filling factor ($f_s < 10^{-4} - 0.5$) reported by \citep{Blustin09}
based on the comparison of the observed radio flux and predicted flux at 1.4 GHz, though that study was focused on objects possessing
an optically thick X-ray absorber.

Therefore, we assume the geometry of the outflowing material to be in the form of a thin ($\Delta R < 1/2~ R$), partially-filled shell, for which we can derive the
total mass $M_{T_i}$ in each kinematic component $i$ by:
\begin{equation}
 M_{T_i}= 4 \pi R_i^2 \Omega \mu m_p N_{H_i},
\end{equation}
where $\mu = 1.4$ is the mean atomic
mass per proton, $m_p$ is the mass of the proton $\Omega$ is the global covering fraction of the outflow, $N_{H_i}$ is the total hydrogen column density for the kinematic component.
In the case of a two-ionization component model we simply have $N_{H_i}=N_{H_{i,lo}}+N_{H_{i,hi}}$. We use $\Omega=0.5$ since outflows are detected in about 50\% of the observed
Seyfert galaxies (e.g.~\citealt{Crenshaw03}). The average mass flow rate $\dot{M}_{T_i}$ is obtained by dividing the total mass of the shell by the dynamical timescale $R_i/v_i$ and
the kinetic luminosity is given by $\dot{E}_{k_i} = 1/2 ~ \dot{M}_{T_i} v_i^2 $:
\begin{eqnarray}
  \dot{M}_{T_i} &=& 4 \pi R_i \Omega \mu m_p N_{H_i} v_i \\
  \dot{E}_{k_i} &=& 2 \pi R_i \Omega \mu m_p N_{H_i} v_i^3. 
  \label{ekeq}
\end{eqnarray}
The advantage of these formulations is that they use of the total hydrogen column density $N_{H_i}$ directly derived from the photoionization modeling of the trough and
thus are independent of the filling factor $f$ of the shell or its $\Delta R$. We list the computed values of $R$, $\dot{M}_T$ and $\dot{E}_k$ for troughs $T2$ and $T3$ in Table~\ref{tabmflo}.

Note that an instantaneous mass flow rate can be defined independently of the dynamical timescale of the outflow by using
the physical definition of $\dot{M}_{T_{i,ins}} = \rho A v$, where $\rho$ is the mass density of the outflowing material
traversing the perpendicular surface $A$ with a velocity $v$. Using the geometry described above this formula simplifies to  $\dot{M}_{T_{i,ins}} = 4 \pi R_i^2 \Omega \mu m_p n_{H_i} f v_i$, where $f$ is the volume filling factor of the shell.
This estimation is directly dependent on the filling factor $f$ of the shell (or its radial extent $\Delta R$), a quantity which is not well constrained observationally.
We however note that, using the definition of the filling factor ($f= N_H/(n_H \Delta R)$), this instantaneous mass flow
rate relates to the average mass flow rate defined in Equation~\ref{ekeq} by the relation $\dot{M}_{T_{i,ins}} = \dot{M}_{T_i}/ (\Delta R / R)$.
Since $\Delta R / R \ltsim 0.1$ (c.f. Section~\ref{distancesect}), this means that the average mass flow
rates, hence the kinetic luminosities, reported in Table~\ref{tabmflo} are lower limits on the instantaneous mass flow rates.

\section{Discussion and conclusions}
\label{discussion}

We analyzed the physical properties of the UV outflow of IRAS F22456-5125 based on high
S/N COS observations. The accurate determination of the column densities associated with the
multitude of ionic species detected in the COS FUV range allowed us to derive the physical 
parameters ($U_H$, $N_H$) of each kinematic components of the outflow. The detection of absorption lines associated with excited states of the low ionization
species \cii\ and \siII\ in two of the kinematic components allowed us to determine the distance to the absorbing
material from the central emission source. In the case of component $T2$, the density diagnostic derived from
the ratio of \siII */\siII\ agrees with that derived from \cii */\cii\ putting the absorbing gas at a distance of
$\sim 10$ kpc. For component $T3$ only the \cii\ diagnostic line is observed suggesting a distance of $\sim 16$ kpc,
though that distance could be underestimated by 30\% in the case of an inhomogeneous absorber model.

The photoionization solutions we find differ from those found by~\citet{Dunn10b} using archival FUSE data(see Table~\ref{pi_solu_table}).
This is due to the limited number of diagnostics available in the FUSE data compared with COS data. The total hydrogen column densities
reported in~\citet{Dunn10b}, derived using only \hi and \ovi\ or \ciii, are generally 0.6 dex higher than the one we find in our analysis. 
Using the ionization timescale (e.g., Krolik and Kriss 1995) along with the assumption that the column density of \hi\ did not change
between two FUSE observations separated by ~21 months, ~\citet{Dunn10b} estimate a lower limit on the distance of $\sim20$ kpc to all
of the kinematic components of the outflow. The distances we find for components T2 and T3 are roughly consistent with this value.
The small discrepancies may be due to the actual lightcurve over the time period being different from simple step-function lightcurve assumed in \citet{Dunn10b}.

Despite the large distance and higher velocity of the outflow compared to the one analyzed in Paper I
the reported kinetic luminosities $\dot{E}_{k_i}$ in Table~\ref{tabmflo} are not energetically significant for AGN feedback.
These scenarios generally require kinetic luminosities to be of the order of a few tenths to a few percent of the Eddington
luminosity $L_{Edd}$ (e.g. \citealt{Scannapieco04}, \citealt{dimatteo05}, \citealt{Hopkins10}) while in the case of IRAS F22456-5125
for which $L_{bol}/L_{Edd} \sim 0.16$ \citep{Dunn08},
we find $\dot{E}_{k_i} \sim 10^{-5} L_{Edd}$. Note that this comparison is probably only a lower limit since it does not
take into account the fact that the outflow probably decelerated and lost energy
through shocks on the way from the launching region to the actual location $R_i$. Moreover, studies of UV outflows in Seyfert galaxies
typically reveal that an associated warm phase
of the outflow, which has an ionization parameter substentially larger than the high ionization phase we report here (see Table~\ref{pi_solu_table}), and usually seen in X-rays \citep{Crenshaw99} can carry 70\%-99\% of the kinetic luminosity of the outflow
(\citealt{Gabel05b}; \citealt{Arav07}). \citet{Dunn10b} analyzed ASCA and XMM-Newton spectra of IRAS-F22456-5125 does not reveal any evidence for
an X-ray warm absorption edge, however, the limited S/N in these data can still allow the presence of a warm phase with significant column density.

Assuming that the gas is in photoionization equilibrium with the central source, we showed that the large number of constraints available
for the determination of the ionization solution of trough $T2$ (and in a weaker way in $T3$) reveals that the
absorbing material can hardly be described by a slab model characterized by a single $U_H$ and $N_H$. Considering
a two ionization solution in which low ionization species are mainly produced in a phase whose ionization
parameter is $\sim$ 1.5 dex smaller than the phase producing the high ionization \nv\ and \ovi, we are able to obtain
a better match to the measured ionic column densities for that component (see Table~\ref{t2__cloudy_model}). If, as inferred from the kinematic
correspondence, the low and high ionization components are located at the same distance from the central
source, we can derive the density of the high ionization component to be $n_{H_{hi}} = (U_{H_{lo}} / U_{H_{hi}}) n_{H_{lo}} \sim
n_{H_{lo}}/30 $.

The observation of several absorption lines corresponding to the \siII\ transition in trough $T2$ allowed us to test
the absorber model by over-constraining the set of fitted parameter by 3 residual flux measurements. For that transition we
find that the PL absorber model describes the observed residual intensities better than the PC model, similar to what was found
by \citet{Arav08} with the modeling of five \feii\ troughs in the spectrum of QSO 2359-1241. Either way, the fits suggest
that only a small fraction of the emission source is covered with optically thick material from that low ionization line. In the same kinematic
component we observe a high covering fraction for the high ionization transition (\civ, \nv, \ovi), almost consistent to a
full covering of the emission source (continuum+BEL+NEL) by those species. Intermediate ionization species like
\cii, \siIII\ or even \siIV\ show clear signs of intermediate covering. These observations suggests a
model where the low ionization phase is formed by relatively small, discrete clumps of denser material
embedded in a lower density, higher ionization phase as suggested by \citet{Hamann98,Gabel05b}.
We however note that the two gas phases are not in pressure equilibrium questioning the survival of
the low ionization clumps in the more homogeneous high ionization phase.

Comparing the properties of the outflow present in IRAS-F22456-5125 and the {\it bona fide} AGN outflow observed in NGC~3783 reveals a more complex situation.
Albeit their similar kinetic luminosity, the in-depth study of the absorption lines presents
in the UV spectrum of NGC~3783 revealed the expected signs of an AGN outflow : line profile variability, high velocities ($v \sim 1400$ km s$^{-1}$), high densities
($n_e \sim 10^{4}$) inferring low distances ($R < 50 pc$) to the central source \citep{Gabel05b}. The large distance, low density, low velocity
material found in IRAS-F22456-5125 is in comparison also typical of galactic winds \citep{Veilleux05}. We reported a similar situation in
Paper I in the case of the outflowing material present in the quasar IRAS F04250-5718. A key question in that case is to determine whether
the galactic wind is driven by the AGN or by starburst activity. While this question has been investigated in the literature,
a definite answer is often out of reach for AGNs in which the sustaining conditions for nuclear activity also favor starburst
activity \citep[e.g.][for a review]{Veilleux05}. While we are not able to determine whether the material is AGN or starburst driven,
the partial covering and the densities higher than the one typically observed in the intergalactic medium deduced from our analysis suggests that the material
is intrinsic to the host galaxy and is hence photoionized by the central source.

We assumed the ionization structure of the outflow is due to radiation from the central
source. We investigated the possibility that the absorber is collisionally ionized by
producing grid-models of $N_H$ versus temperature with a fixed ionization parameter of
$10^{-5}$. At temperatures $\sim 10^5$~K, we reproduce all the metal lines except  \nv\ and
\ovi\ in trough T3. By including another, hotter phase ($T \approx 10^{5.5}$~K),
all of the metal lines are reproduced. However, \hi\ is underpredicted by high temperature models
(a factor of $\sim$100 for $\sim 10^{4.8}$~K at $\log{N_H} = 19$). We are therefore lead to the conclusion that the ionization structure of the absorber
is dominated by photoionization.

\section*{ACKNOWLEDGMENTS}
B.B. would like to thank G. Schneider and B. Stobie for providing their IDL implementation of the IRAF RL algorithm, and
G. Letawe for useful discussions. B.B. thanks also S. Penton for the introduction to the HST/COS pipeline. We thank the
anonymous referee for a careful reading of the manuscript and suggestions that helped improve the paper. We acknowledge
support from NASA STScI grants GO 11686 and GO 12022 as well as NSF grant AST 0837880.


\bibliographystyle{apj}

\bibliography{astro}{}


\end{document}